\documentclass[3p,times,twocolumn]{elsarticle}
\usepackage{ecrc-ac}
\usepackage{hyperref}
\hypersetup{colorlinks=true, linkcolor=blue, filecolor=blue, urlcolor=red}
\volume{00}
\firstpage{1}
\journalname{Nuclear Physics B Proceedings Supplement}
\runauth{T. Riemann}
\jid{nuphbp}
\jnltitlelogo{Nuclear Physics B Proceedings Supplement}
\usepackage{amssymb}
\usepackage[figuresright]{rotating}
\newcommand{\be}{\begin{equation}}
\newcommand{\ee}{\end{equation}}
\newcommand{\beq}{\begin{equation*} }
\newcommand{\eeq}{\end{equation*}  }
\newcommand{\bqa}{\begin{eqnarray*} }
\newcommand{\eqa}{\end{eqnarray*}}
\newcommand{\bea}{\begin{eqnarray}}
\newcommand{\eea}{\end{eqnarray}}

\newcommand{\nn}{\nonumber}

\begin{document}

\begin{frontmatter}

\dochead{}
\title{
\vspace*{-4.5cm}\flushleft{\begin{small}         \tt SFB/CPP 14-97            \end{small}}
\\[4cm]
Massive Feynman integrals and electroweak corrections
}

\author[label1]{Janusz Gluza}
\author[label2]{Tord Riemann}

\address[label1]{Institute of Physics, University of Silesia, Uniwersytecka 4, 40007 Katowice, Poland}
\address[label2]{15711 K{\"o}nigs Wusterhausen, Germany}

\begin{abstract}
%
There are steady advances in the calculation of electroweak corrections to massive scattering problems at colliders, from the 
very beginning in the nineteen seventies until contemporary developments.
Recent years brought a remarkable progress due to new calculational technologies.
This was motivated by demands from phenomenological applications at particle accelerators: higher multiplicities of the final states, 
extreme kinematics, need of higher precision and thus of higher orders in perturbation theory.  
We describe selected contributions from the project ``Massive particle production'' of Sonderforschungsbereich/Transregio 9 of Deutsche 
Forschungsgemeinschaft. 
\end{abstract}


\end{frontmatter}

\section{Introduction}
\label{sec:1}
The calculation of observable quantities for high energy colliders became more and more involved in recent years, although the basic 
understanding of perturbative quantum field theory has been settled decades ago.
The term ``calculation'' has two sides here to be taken into account, of quite different origin.
First, one has to derive formulae for the quantity of interest, with a sufficient accuracy in order to match experimental needs. This is 
part of theoretical work in the classical understanding.
But, by time the answers get more involved, both in quantity and in complexity. Also, the singularity behaviour becomes worse.
As a consequence, the result of theoretical research to be disseminated is often not only an analytical formula written in an article, but 
also a piece of more or less sophisticated software.
This is fine, but it raises new questions of cooperation.
Software has to be supported in a rapidly developing world of computing.
How to distribute software in an appropriate manner, thereby 
respecting the authors' rights in a satisfactory way, but at the same time not 
too much hindering its use?
 Let us remind that software use in nearly all realistic cases means also adaptation and so changing the original creation. 

Since we are working in the field of particle phenomenology since decades, we collected some experience with all these aspects, to some 
extent we even 
contributed to the culture of practicioning.
We come back to the point in section ~\ref{sec-dissemination}.

In sections ~\ref{sec-zfitter} to ~\ref{sec-ambre} we survey part of research performed in the research group B1 ``Massive particle 
production'' of Sonderforschungsbereich/Transregio 9 of 
Deutsche Forschungsgemeinschaft.
Due to the calculational difficulties it took few years after the concise formulation of the perturbative renormalization of the 
electroweak theory by t' Hooft and Veltman  
\cite{'tHooft:1972fi,'tHooft:1972ue} and after the invention of SCHOONSCHIP \cite{Strubbe:1974vj}.
A famous piece of work was Veltman's study of the $\rho$ parameter with the observation that high particle masses may show up at low 
energy \cite{Veltman:1977kh}.
First detailed studies of the calculational techniques and 
of the consequences for phenomenology came out soon, notably \cite{'tHooft:1978xw,Passarino:1978jh}.
Since then, much effort has been concentrated to the refinement of predictions of perturbative effects in the Standard Model and beyond.

Calculations have been done for many quantities, notably the weak corrections to the $Z$ boson parameters $\rho_Z$ and 
$\sin^{2,eff}_W$; at one loop e.g. in \cite{Bardin:1980fe,Bardin:1981sv,Akhundov:1985cf,Akhundov:1985fc}, and later also with higher order 
corrections predicted 
by the  electroweak theory and by QCD \cite{Bardin:1997xq,Awramik:2004ge,Awramik:2006uz,Freitas:2012sy,Freitas:2013dpa}.
These higher-order calculations have to be performed, but they have also to be inserted into phenomenological tools.

Although a lot of the material presented here is applied also to  LHC physics, we will concentrate on higher-order contributions to $e^+e^-$ 
annihilation, mainly arising from loop corrections:
\bea
\label{eq-2f}
e^+e^- \to f^+f^-, ~~ f^+f^-(\gamma), ~~f^+f^-\gamma, ~~ f^+f^-\gamma(\gamma).
\eea
A large part of the present study is devoted to  the treatment of single Feynman integrals.
They are the building blocks of Feynman diagrams related to some observable.
  We will consider an arbitrary $L$--loop integral $G(X)$ with loop momenta $k_l$ , with $E$ external
legs with momenta $p_e$  and with $N$ internal lines with masses $m_i$ and propagators $1/D_i$: 
\be \label{eq-2}
 G(X) = \frac{1}{(i \pi^{d/2})^L} \int 
 \frac{d^d k_1 \ldots d^d k_L \, X(k_1, \ldots, k_L)}
      {D_1^{n_1} \ldots D_i^{n_i} \ldots D_N^{n_N}}   ,
\ee
with
\bea
 d &=& 4 - 2 \epsilon ,
\\
 D_i &=& q_i^2 - m_i^2 = \left[
 \sum \limits_{l=1}^{L} c_i^l k_l + \sum \limits_{e=1}^{M} d_i^e p_e
 \right] - m_i^2 ,
\eea
where
$X(k_1, \ldots, k_L)$ stands for tensors in the loop momenta.

\section{ZFITTER\label{sec-zfitter}}
ZFITTER \cite{Bardin:1999yd,Arbuzov:2005ma} is a long-term project, dating back to the nineteen seventies. 
The aim is a state-of-the-art description of 
\bea
\label{eengamma}
e^+e^- \to (\gamma, Z) \to f^+f^-(n\gamma)
\eea
in the Standard Model.
A description of the project has been published quite recently  
\cite{Akhundov:2014era}.
Since 1989 ZFITTER is among the standard software packages for the description of the  $Z$ boson resonance at LEP.
Further, it was used for predictions of the top-quark 
and Higgs-boson masses from radiative corrections in the Standard Model prior to their discoveries.
Until about 1992, ZFITTER rested mainly on theoretical work done by its authors on complete one-loop electroweak corrections in the 
Standard Model. In the nineteen nineties it become more and more important to integrate higher-order corrections derived by other authors 
and to support the users from experimental groups, notably from DELPHI, L3, OPAL, and also from the LEPEWWG. 
This is documented in the ``LEP electroweak working group report'' of 1995 \cite{Bardin:1997xq} and references therein.
The seminal review studies of (\ref{eengamma}) by the LEP community for LEP 1 in 2005 \cite{ALEPH:2005ab} and LEP 2 in 2013 
\cite{Schael:2013ita} rest to a large extent on ZFITTER v.6.42 \cite{Bardin:1999yd,Arbuzov:2005ma}.  
ZFITTER became the ``etalon'' software for the $Z$-boson resonance studied for many years at LEP 1 und at LEP 2.
Among the main results of LEP are the following, quoted from the ``Review of Particle Physics'' (2012) \cite{Beringer:1900zz}:
\bea
M_Z &=& 91.1876 \pm 0.0021 {\rm~~GeV},
\nonumber \\
\Gamma_Z  &=& 2.4952 \pm 0.0023 {\rm~~GeV},
\nonumber \\
\sin^2 \theta_{\rm weak} &=& 0.22296 \pm 0.00028 ,
\nonumber \\
\label{sinw2-lep}
\sin^2 \theta_{\rm lept}^{\rm eff} &=& 0.23146 \pm  0.00012,
\nonumber \\
\sin^2 \theta_{Z}^{\rm MS} &=&  0.23116 \pm 0.00012,
\nonumber \\
 N_\nu &=& 2.989 \pm 0.007.
\eea

\begin{figure}[h]
\hspace*{-1.cm}
\center{\includegraphics[width=7.5cm]{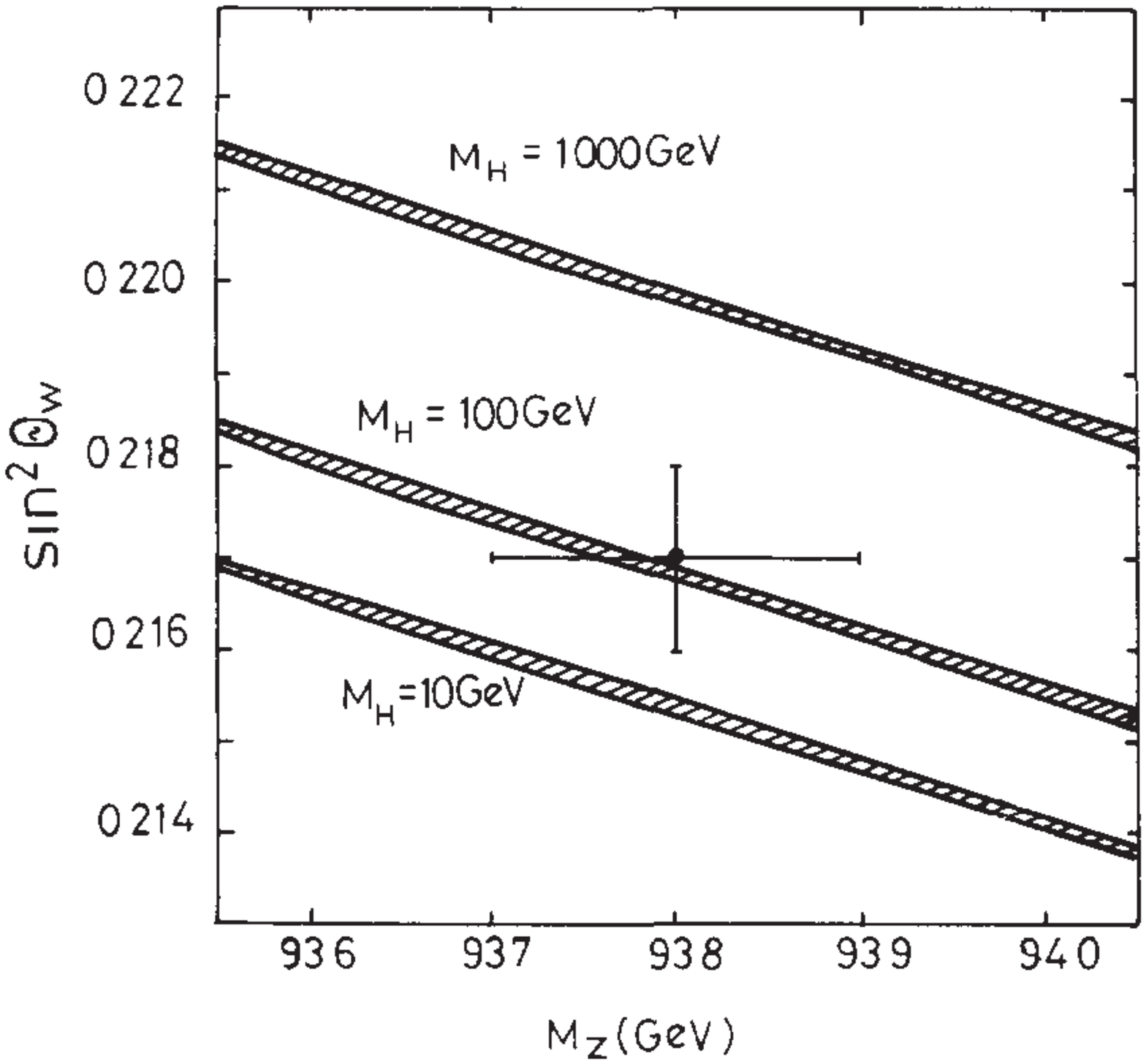}}
\caption{
\label{plb166fig1}
The first ever plotted LEP observables' dependence on the Higgs mass in
the Standard Model (reprinted from Physics Letters, A.~Akhundov, D.~Bardin, and T.~Riemann, ``Hunting the
hidden standard Higgs'', volume B166, p. 111, Copyright (1986) \cite{Akhundov:1985cf}, with permission from
Elsevier.)
Graph of $\sin^2\theta_W$ versus $M_Z$, influenced by $M_H$ through radiative corrections.
 The thickness corresponds to the range 30 GeV $< m_t < $ 40 GeV, the error bars indicate the accuracy expected at Z boson factories.
}
\end{figure}

\begin{figure}[h]
\vspace*{-0.3cm}
\center{\includegraphics[angle=0,width=5.5cm]{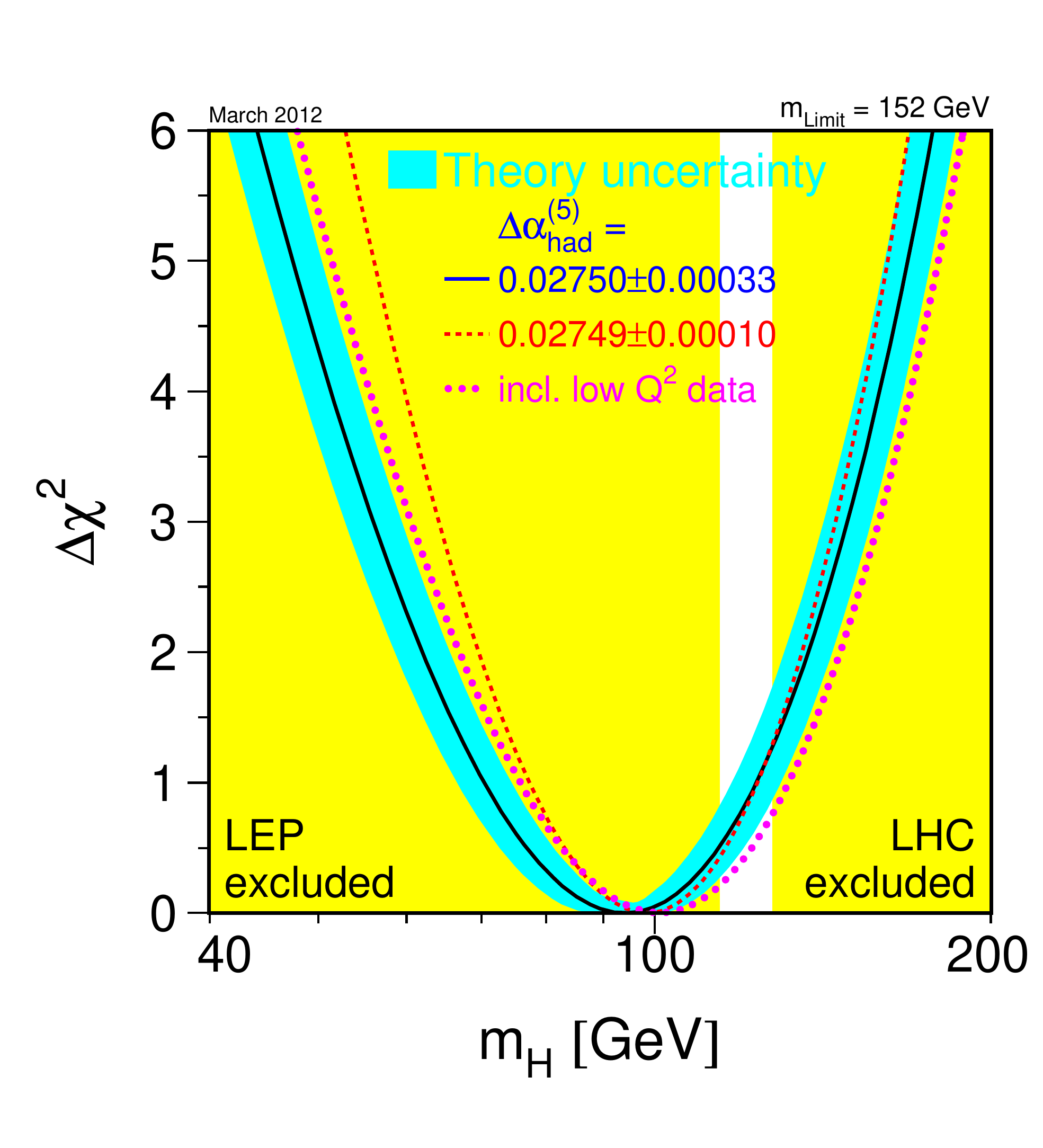}}
\vspace*{-0.3cm}\caption[Higgs mass and LEP measurements]{
Blue-band plot of the LEPEWWG
\cite{lepewwg-webpage-jan-2013} with a Standard Model Higgs boson mass prediction based on combined world
data from precision electroweak measurements.
}
\label{w12_blueband}
\end{figure}

\begin{figure}[h]
\begin{center}
\vspace*{-0.1cm}
\includegraphics[width=6.cm]{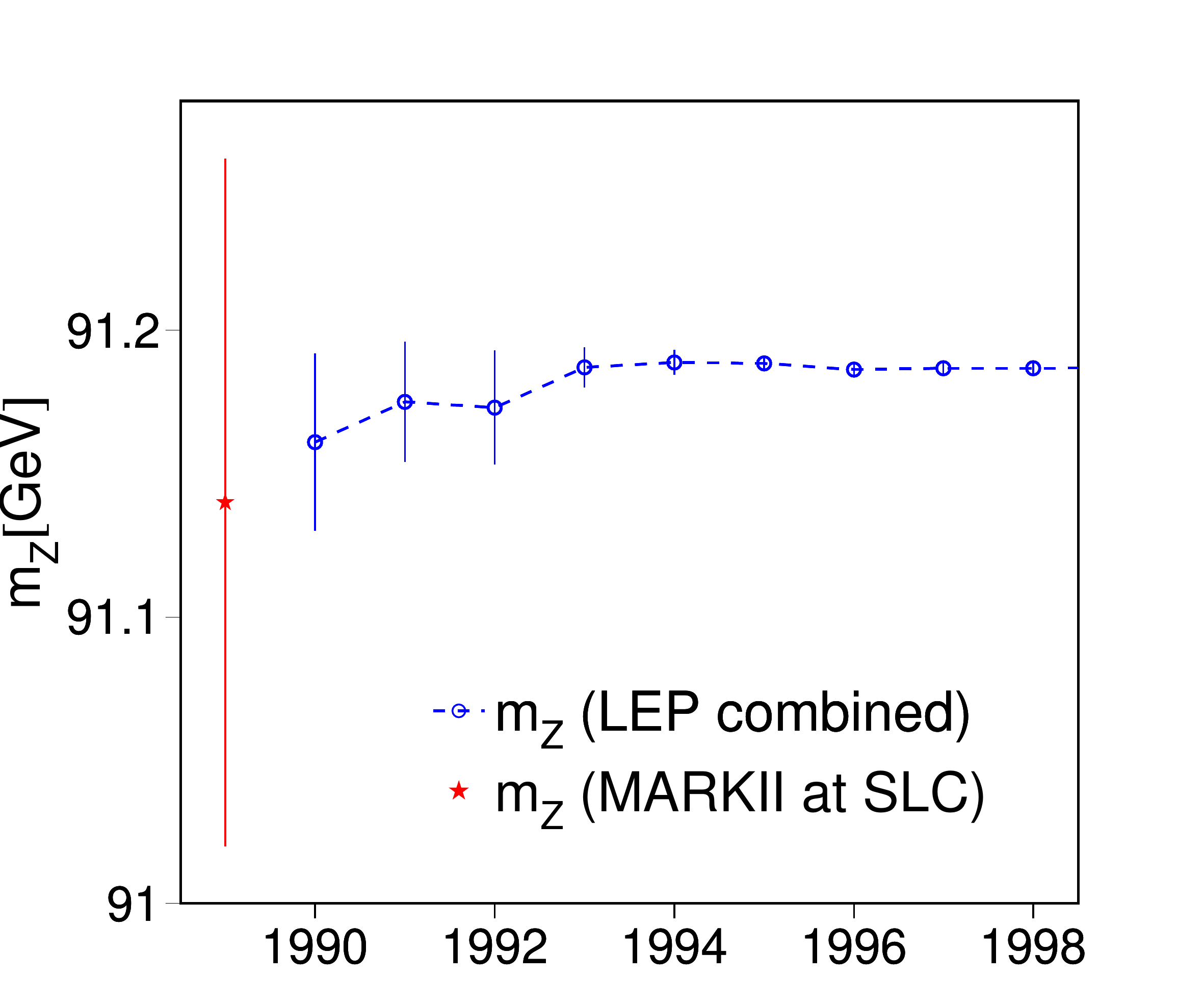}
\end{center}
\vspace*{-0.5cm}
 \caption{
$Z$ boson  mass measurements at LEP. Earlier measurements are from UA1, UA2 at SPS (CERN) (see
text, not shown in plot) and from MARKII at SLC (SLAC). Reprinted from \cite{Akhundov:2014era}, with permission from
Springer Verlag under licence number 3494820307523.}
\label{fz1}
\end{figure}

\begin{figure}[h]
\begin{center}
\includegraphics[width=6.cm]{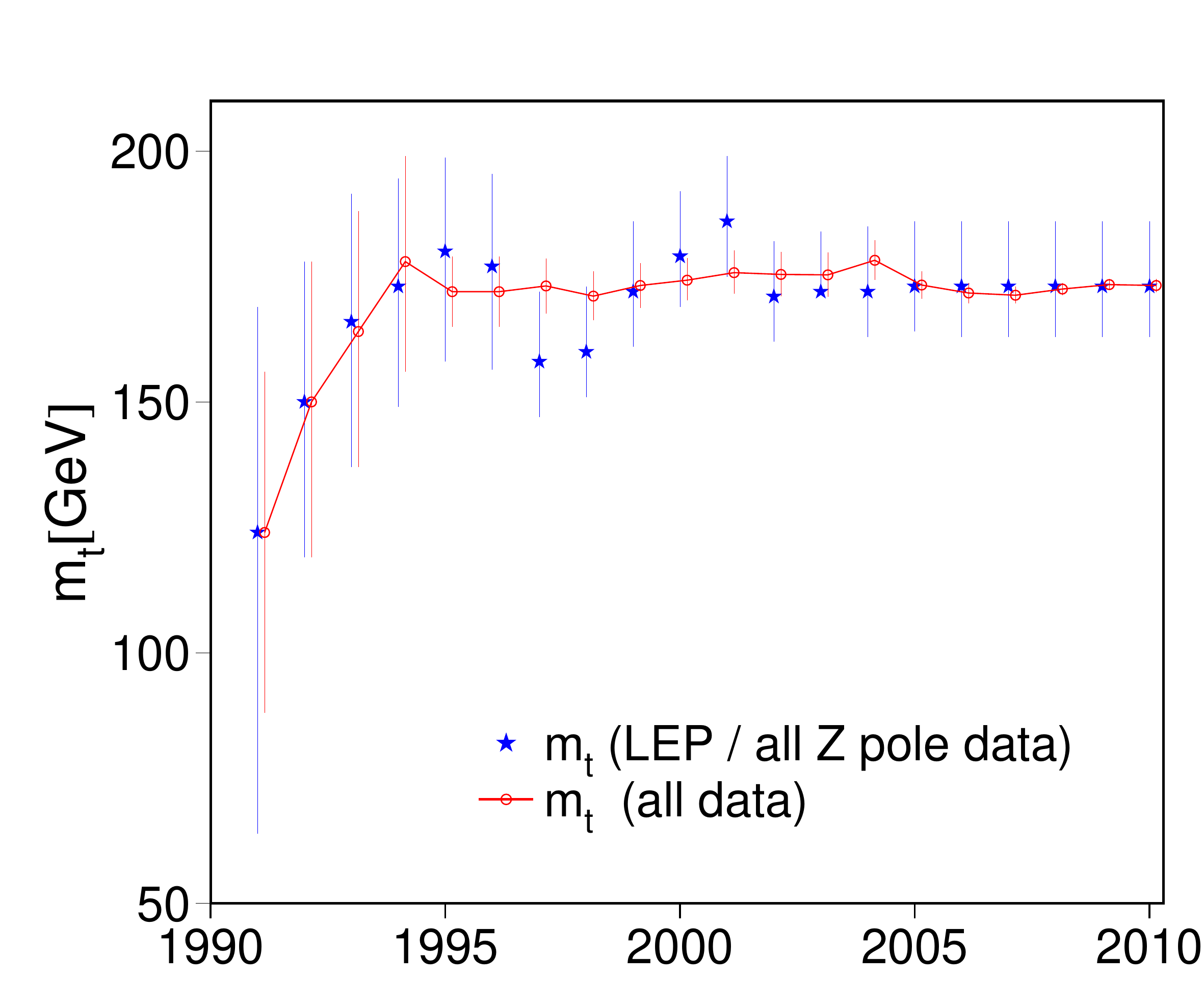}
\end{center}
\vspace*{-0.5cm}
 \caption{Top quark mass measurements. Reprinted from \cite{Akhundov:2014era}, with permission from
Springer Verlag under licence number 3494820307523.}
\label{fz2}
\end{figure}

\begin{figure}[h]
\begin{center}
\includegraphics[width=6.cm]{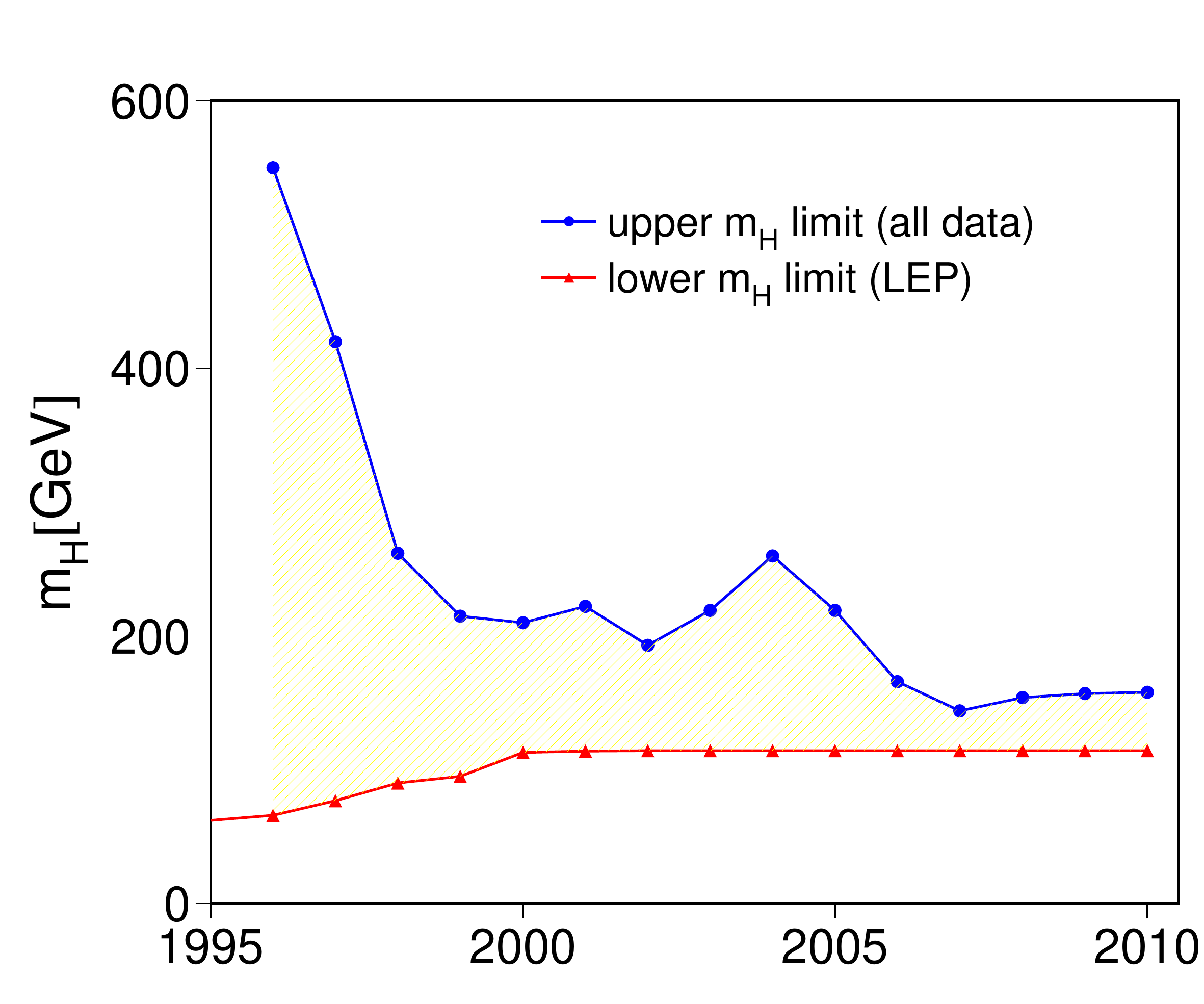}
\end{center}
\vspace*{-0.5cm}
 \caption{
Higgs boson  mass measurements. The upper limits and the fit values for $M_H$ derive from a
combination of virtual corrections to LEP and similar data, top and $W$ mass measurements, performed by the
LEPEWWG. The lower mass limit is due to LEP direct searches. The lower limits from data combinations are not
shown. Reprinted from \cite{Akhundov:2014era}, with permission from Springer Verlag under licence number 3494820307523.
}\label{fz3}
\end{figure}

A similar analysis, also based on ZFITTER, has been published by ALEPH, DELPHI, L3, OPAL, LEPEWWG
in 2013
\cite{Schael:2013ita}.
The constraint 
\bea
m_t = 178^{+11}_{-8} ~ {\rm GeV} 
\eea
is obtained from the virtual corrections, in good agreement with the much more
precise direct measurement of about { $m_t$} $= 173.2 \pm 1 $ GeV \cite{LissRPP:2014}.
For the Higgs boson mass, they predict:
\bea \nonumber
 M_H &=& 118^{+203}_{-64}  ~ {\rm GeV,~~only~~ LEP},
 \\ \nonumber
M_H  &=& 122^{+59}_{-41}  ~  {\rm GeV,~~plus~~} m_{t},
\\ \nonumber
M_H&=& 148^{+237}_{-81}  ~ {\rm GeV, ~~ plus~~} M_W, ~\Gamma_W,
\\
 M_H &=& 94^{+29}_{-24} ~  {\rm GeV, ~~plus~~} m_t,~ M_W,~ \Gamma_W.
\eea
An update is \cite{ErlerFreitasRPP:2014}, where it is quoted $M_H =  89^{+22}_{-18}$ GeV, or $M_H < 127$ GeV (90\% c.l.).   
In 2012, the LHC collaborations discovered a scalar particle with a mass of about 125
GeV \cite{atlas:2012gk,CMS:2012gu}. The present best value is $M_H = 125.6 \pm 0.3$ GeV \cite{CarenaRPP:2014}.
This might be illustrated by the famous blue band plot of the LEPEWWG \cite{lepewwg-webpage-jan-2013,nobel-higgs:2013}, which we reproduce 
in figure \ref{w12_blueband}, together with the presumably first proposal of an electroweak precision plot in figure ~\ref{plb166fig1}. 
The development of precision predictions is nicely illustrated in figures~\ref{fz1} to ~\ref{fz3}. 

\begin{figure}[h]
\begin{center}
\vspace*{-0.5cm}
\hspace*{-0.7cm}
\includegraphics[width=9.5cm]{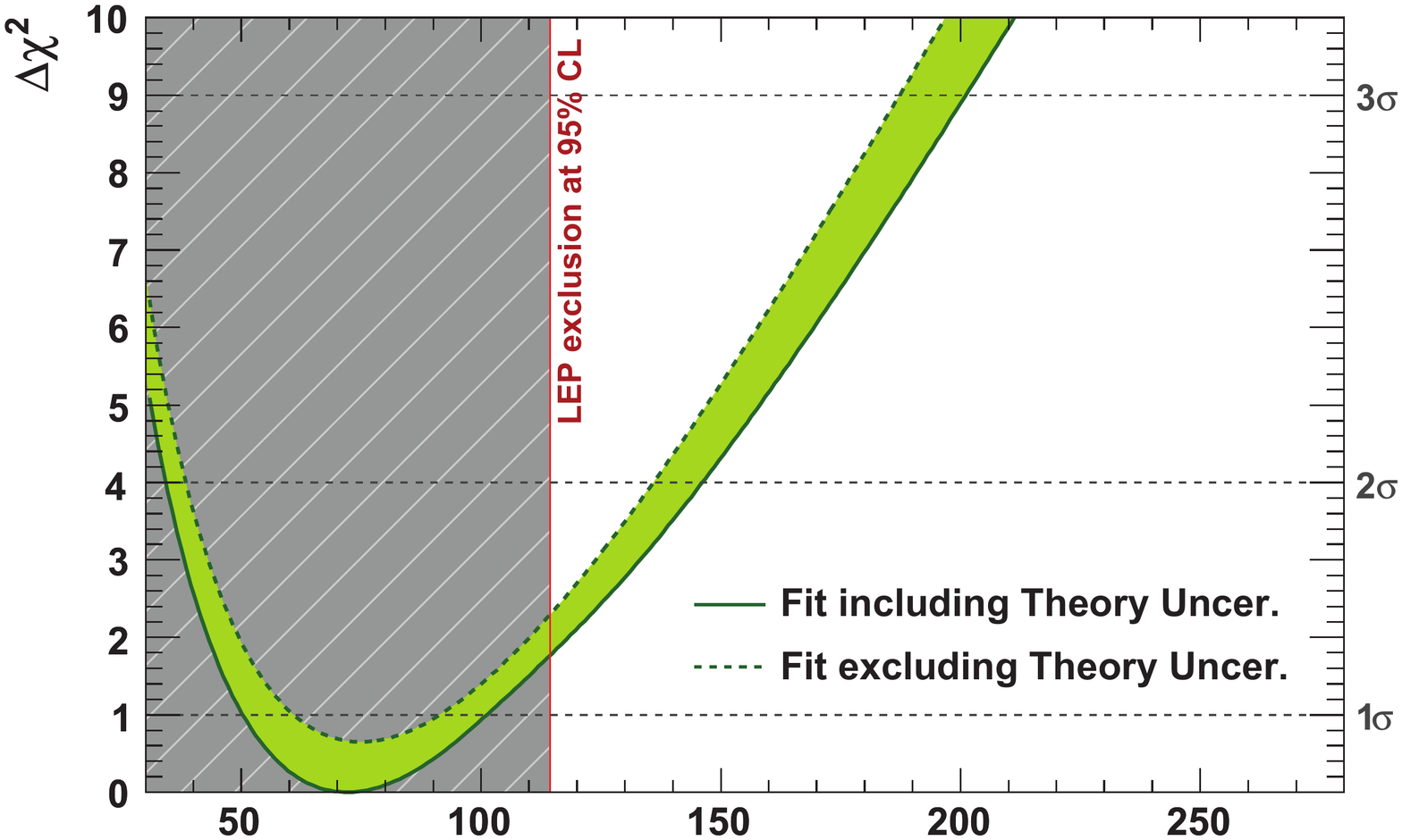}
\end{center}
\vspace*{-1.2cm}
\caption{Gfitter plot on the Higgs boson mass, reproduced from figure 6.8 of \cite{Goebel:2008}.
\label{gf2008}
}
\end{figure}

It is pointed out in \cite{Schael:2013ita} that there are, besides ZFITTER, two alternative approaches for precision Standard Model 
tests available. One approach is practiced in the ``Review of particle physics'' of the Particle Data Group \cite{Nakamura:2010zzi}, which 
traces to a large extent back to ZFITTER. 
The second approach is the Gfitter project. In fact, at the webpage 
\href{http://gfitter.desy.de/}{http://gfitter.desy.de/} one finds a lot of data similar to the blue band plot of the LEPEWWG, figure 
~\ref{w12_blueband}. We reproduce here as an example the figure ~\ref{gf2008}.  
 The Gfitter plots are not due to independent calculations.  
They have been made with the software Gfitter/gsm  \cite{Goebel:2008} which originates from the Standard Model library of ZFITTER v.6.42 
\cite{Bardin:1999yd,Arbuzov:2005ma}, see 
\href{http://zfitter-gfitter.desy.de/}{http://zfitter-gfitter.desy.de/}.\footnote{The authors of 
Gfitter/gsm are M. Goebel, J. Haller, A. Hoecker.}
 Like the diploma thesis \cite{Goebel:2008}, all the Gfitter publications, talks and proceedings contributions from December 2007 until 
July 2011 make use of Gfitter/gsm, see 
webpage \href{http://fh.desy.de/projekte/gfitter01/Gfitter01.htm}{http://fh.desy.de/projekte/gfitter01/Gfitter01.htm}. 

So we are faced with the unfortunate situation that for a large extent of precision predictions in the Standard Model there is basically 
only ZFITTER as a supported tool with reproducible features.\footnote{We like to mention that \cite{Ciuchini:2013pca} is an 
independent programming of ZFITTER's Standard Model library's physics contents. We are grateful to S. Mishima for indicating to us few 
typos in ZFITTER v.6.42; fortunately, they are of negligible numerical influence.}  

ZFITTER is among the software tools with longest history of open access in the particle physics community. The program traces back to 
times when 
there was no public internet, and the files were exchanged by floppy disks. 
We estimate the human investment into ZFITTER as follows:
\begin{itemize}
 \item 
{ 2.2 million Euro}
\\
derived from {30 years FTE} (staff researcher full time equivalent with 74,000 Euro per FTE)
 \item 
{ 1.1 million Euro}
\\
1/2 of the total amount for project management, publications, numerical tests, user support etc.
 \item 
{ 1.1 million Euro}
\\
1/2 of the total amount for software  = QED corrections + Standard Model library, resulting in:
\\
{ 550,000 Euro} $\to$ QED corrections
\\
{ 550,000 Euro} $\to$ { Standard Model library}
\end{itemize}

The ZFITTER webpage with lots of information on older versions of ZFITTER is
\href{http://zfitter.education}{http://zfitter.education}.
ZFITTER v.6.42 is exclusively available from  
\\
\href{http://cpc.cs.qub.ac.uk/summaries/ADMJ}{http://cpc.cs.qub.ac.uk/summaries/ADMJ}. A 
download is possible only after active agreement on the licence conditions shown in the pop-up window ``CPC LICENCE ALERT''.

In recent years, new regions of application have been explored.
There are quite interesting analyses of the Drell-Yan cross section by the CDF collaboration \cite{Aaltonen:2013wcp,Aaltonen:2014loa}; see 
also the CMS study \cite{Chatrchyan:2011ya}. 
The high luminosity data expected at BELLE 2 will also deserve the application of ZFITTER; see 
also section \ref{sec-topfit}. 
In many cases, a support by the authors, 
including theoretical adaptations, has been welcomed by the experimentalists.  

The theoretical description of the $Z$ resonance measurements at a future Giga-$Z$ factory  will deserve electroweak two-loop precision.
Here, certain multi-scale two-loop vertex-type Feynman integrals are hard to calculate, but seem feasible with present technology. 
Once 
this is done, the complete electroweak two-loop corrections to the $Z$ resonance will be known.   
For further remarks on this subject see section \ref{sec-ambre}.

\section{TOPFIT\label{sec-topfit}}
\begin{figure}[t]
\vspace*{0.4cm}
\hspace*{-1.2cm}
\includegraphics[width=9.0cm]{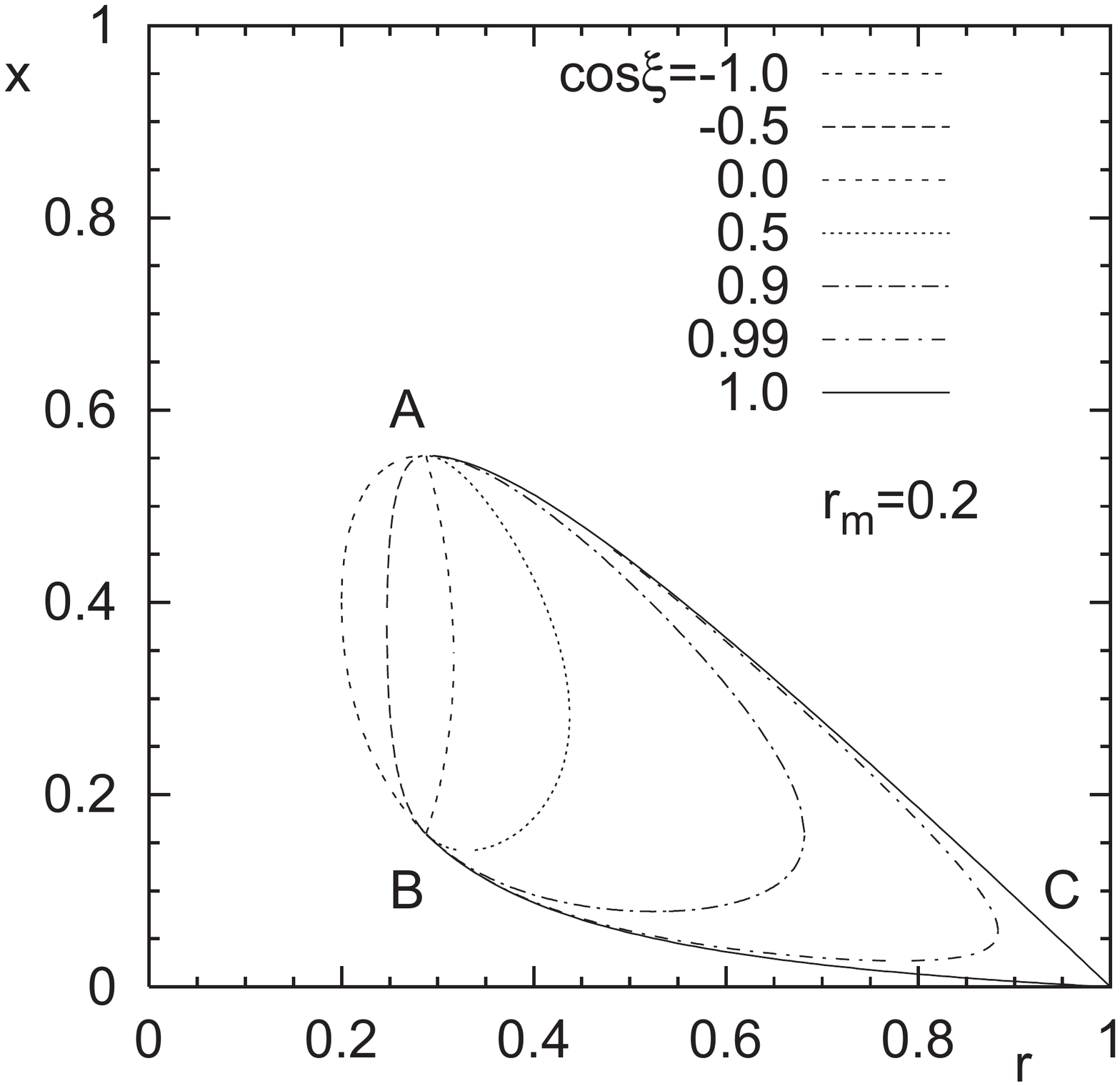}
\caption{\label{fig-acol}
Kinematic region of $r$ and $x$ for different values of the acollinearity
angle $\xi$ for 
$r_m = 4m_f^2/s$ = 0.2. Figure from \cite{Fleischer:2002kg}.
}
\end{figure}

ZFITTER assumes the external fermions in (\ref{eq-2f}) and (\ref{eengamma}) to be light.
In $\mu$ and $\tau$ pair production at meson factories potentially, and naturally in top pair 
production in the continuum at the ILC, the final state mass is essential.
  For the description of complete electroweak corrections with exact account of the final state mass, one may use  TOPFIT 
\cite{FRW:2002sw}.
The virtual corrections and a semi-analytical treatment of hard photonic corrections with certain experimental cuts has been described in 
\cite{Fleischer:2003kk}.   
As an example of the sophisticated situation we reproduce a Dalitz plot with an acollinearity cut in figure \ref{fig-acol}.
The variables are: $\xi=\pi-\sphericalangle(f^- f^+)$, $x =2p(\gamma)p(f^+)/s$, and $r = 1-2E(\gamma)/\sqrt{s}$.

Phenomenological applications were worked out in 
\cite{Fleischer:2002rn,Fleischer:2002kg,Fleischer:2002nn11,Accomando:2004sz,Fleischer:2003aa,Hahn:2003ab}.
The Born cross section depends on the running electromagnetic coupling $\alpha_{em}$ and, for massless fermion pair production, on four  
electroweak form factors $\rho_Z, \sin^{2,eff}_{e}, \sin^{2,eff}_{f}, \sin^{2,eff}_{ef}$. Final state mass adds two degrees of freedom:
\begin{eqnarray}
  \label{eq:sigma6f}
\frac{d\sigma_{Born}}{d\cos\theta} 
=
\frac{\pi \alpha^2}{2s} 
~c_f~ 
\beta
~2 \Re e \Bigl[ 
  (u^2+t^2+2m_{f}^2s)
\nonumber
\\
\left(\bar{F}_1^{11} \bar{F}_1^{11,B*} +\bar{F}_1^{51} \bar{F}_1^{51,B*} \right)
\nonumber
\\
+~
  (u^2+t^2-2m_{f}^2s)
\left(\bar{F}_1^{15} \bar{F}_1^{15,B*} +\bar{F}_1^{55} \bar{F}_1^{55,B*} \right)
\nonumber
\\
+~  { (u^2-t^2)}
\bigl( \bar{F}_1^{55} \bar{F}_1^{11,B*}+ \bar{F}_1^{15} \bar{F}_1^{15,B*}
  + \bar{F}_1^{51} \bar{F}_1^{51,B*}   
\nonumber
\\
 + \bar{F}_1^{11} \bar{F}_1^{55,B*} \bigr)
\nonumber
\\  
+~ 2m_{f}(tu-m_{f}^4) \left( \bar{F}_3^{11}  \bar{F}_1^{11,B*} 
                       +\bar{F}_3^{51}  \bar{F}_1^{51,B*} \right)
\Bigr]\,.
\end{eqnarray}
The $s,t,u$ are the Mandelstam variables, $c_f$ the color factor, and the $F, \bar F$ are form factors \cite{Fleischer:2003kk}.
The project was originally devoted to physics at the ILC, and several phenomenological studies were performed for the corresponding energy 
range \cite{Hahn:2003ab,Fleischer:2002nn11}.
In tables \ref{table_tau_500}  for $\tau$ leptons and \ref{tab-topfitgrace} for top quarks we reproduce comparisons of differential cross 
sections    and in table  \ref{tab-topfitgrace2} integrated observables with hard photons included for top quark production.
The tables demonstrate the impressive accuracy level achieved by comparing the numerics of different groups. 

\newcommand{\red}{\color{Red}}
\newcommand{\green}{\color{Blue}}
\newcommand{\black}{\color{Black}}

\def \oa{$\mathcal{O}(\alpha)$~}
\def \Oa{\mathcal{O}(\alpha)}
\def \Oatwo{\mathcal{O}(\alpha^2)}
\def \Oathree{\mathcal{O}(\alpha^3)}
\def \Oafour{\mathcal{O}(\alpha^4)}
\def \Mcal{\mathcal{M}}
\newcommand \Mcalf[2]{\mathcal{M}^{(#1)}_{#2}}
\newcommand \Mcals[2]{{\mathcal{M}^{(#1)}_{#2}}^*}
\def \bM{\mathbf{M}}
\def \bF{\mathbf{F}}
\def \ct{\cos{\theta}}
\def \emo{\cdot{} 10^{-1}}
\def \emt{\cdot{} 10^{-2}}

\begin{table}[b!]
{\footnotesize
$$
  \begin{array}{|r|l|l|}
  \hline
  \multicolumn{3}{|c|}{\vphantom{\Big|}e^+e^-\to \tau^+\tau^-\qquad \sqrt s = {500 GeV}} \\
  \hline\vphantom{\Big|}
  \ct &
  \left[\frac{d\sigma}{d\cos\theta}\right]_{\textrm{Born}} 
&
  \left[\frac{d\sigma}{d\cos\theta}\right]_{\textrm{B+w+QED+soft}} 
\\
  \hline\hline
-0.9 & 0.094591~02171~8632{\bf9}  
& 0.092419~02671~1{\bf4061}  
\\
-0.9 & 0.094591~02171~8632{\bf7}  
& 0.092419~02671~1{\bf8656}  
\\
\hline
-0.5 & 0.089298~53117~7985{\bf8}  
& 0.086699~48248~6{\bf5248}  
\\
-0.5 & 0.089298~53117~7985{\bf6}  
& 0.086699~48248~6{\bf9477}  
\\
\hline
 0.0 & 0.15032~16827~75192 
& 0.14359~79492~086{\bf48} 
\\
 0.0 & 0.15032~16827~75192 
& 0.14359~79492~086{\bf18} 
\\
\hline
 0.5 & 0.28649~90174~53525 
& 0.28258~86777~59{\bf811} 
\\
 0.5 & 0.28649~90174~53525 
& 0.28258~86777~59{\bf161} 
\\
\hline
 0.9 & 0.44955~18970~14604 
& 0.47648~29191~20{\bf038} 
\\
 0.9 & 0.44955~18970~14604 
& 0.47648~29191~19{\bf623} 
\\
\hline
\end{array}
$$
}
\caption{Differential cross-sections in picobarn for selected scattering angles
 for $\tau$-production at $\sqrt{s} = 500$ GeV.  The three columns contain 
 the Born cross-section, Born including only the weak $O(\alpha)$ corrections, 
 and Born including the weak and soft photonic $O(\alpha)$ corrections.  For each 
 angle, the first row represents the TOPFIT result of the Zeuthen group 
 while the second contains the Feynarts/Feyncalc calculation of the Munich 
 group. Table shortened from SFB/CPP-03-13 \cite{Hahn:2003ab}.}
\label{table_tau_500}
\end{table}

\normalsize

\def\sig{\left[\frac{\displaystyle{\mathrm{d}\sigma}}{\displaystyle{\mathrm{d}\cos \, \theta}}\right]}

\begin{table}[h]
{
\footnotesize
$$
\begin{array}{|r|l|l|l|}
\hline 
\vrule height 3ex depth 0ex width 0ex
\cos\theta  
& \sig_{{Born}}  
& \sig_{{SM}}
&  \sig_{{tot}}
\\ 
\hline 
\hline  
-0.9  
&  0.108839194075   
& +0.11408410 & 0.13144
\\
& 0.108839194075    
& -0.00205485\bf{8} & 0.132{\bf{29}}
\\[-0.2mm]
&  0.108839194076   
& -0.00205485\bf{9}  & 0.132\bf{06}(12)
\\
\hline 
-0.5  
&  0.142275069392   
& +0.14308121  & 0.15973
\\
&   0.142275069392  
& -0.01512903\bf{8} & 0.160\bf{29}
\\
&  0.142275069393   
& -0.01512903\bf{9}    & 0.160\bf{13}(13)
\\
\hline 
+0.0 
&  0.225470464033   
& +0.21718801  & 0.23638
\\
&   0.225470464033  
& -0.04321416\bf{9} & 0.23\bf{476}
\\
&  0.225470464033   
& -0.04321416\bf{8}   & 0.23\bf{513}(14) 
\\
\hline 
+0.5  
&  0.354666470332   
& +0.32933727  & 0.35651
\\
&   0.354666470332  
& -0.09550125\bf{7} & 0.35\bf{062}
\\
&  0.354666470332   
& -0.09550125\bf{2} & 0.35\bf{104}(17)
\\\hline 
+0.9  
&  0.491143715767   
& +0.44290816  & 0.48796
\\
&  0.491143715767   
& -0.16747886  & 0.477\bf{68}
\\
&  0.491143715767   
&  -0.16747886  & 0.477\bf{09}(21)
\\ \hline
\end{array}
$$
\caption{Various differential cross sections, last column includes hard photon emission.
  The upper and lower rows correspond to the {TOPFIT} 
  and GRACE
  approach, respectively, $\sqrt{s}$ = 500 GeV.
Table shortened from \cite{Fleischer:2002nn11}.
\label{tab-topfitgrace}
}
} 
\end{table}


\begin{table}[h]
{\footnotesize 
$$
\begin{array}{|r|c|c|c|c|c|c|}
\hline 
\vrule height 3ex depth 0ex width 0ex
 \sqrt{s}~~    &\sigma_{\mathrm{tot}}^0 & A_{\mathrm{FB}}^0 
& \sigma_{\mathrm{tot}} & A_{\mathrm{FB}}
\\ 
\hline 
\hline  
500 & \mathrm{T:}~ 0.51227\bf{44} & 0.41460\bf{39}  
&0.5263\bf{37}  &  
0.36\bf{2929}
\\ 
    & \mathrm{G:}~ 0.51227\bf{51} & 0.41460\bf{42}  
& 0.5263\bf{71} & 
0.36\bf{3140}
\\
\hline
1000 &\mathrm{T:}~  0.155918\bf{5} & 0.56417\bf{06}  
&
0.1719\bf{16}& 0.4888\bf{69}
\\
   & \mathrm{G:}~  0.155918\bf{7} & 0.56417\bf{10}  
& 0.1719\bf{31} & 
0.4888\bf{72}
\\ 
\hline
\end{array}
$$
}
\caption{Total cross sections in pbarn and forward-backward asymmetries. Corrections include hard photon emission.
Table shortened from \cite{Fleischer:2002nn11}.
\label{tab-topfitgrace2}
}
\end{table}

\begin{figure}[h]
\hspace*{-0.2cm}
\includegraphics[width=8.0cm]{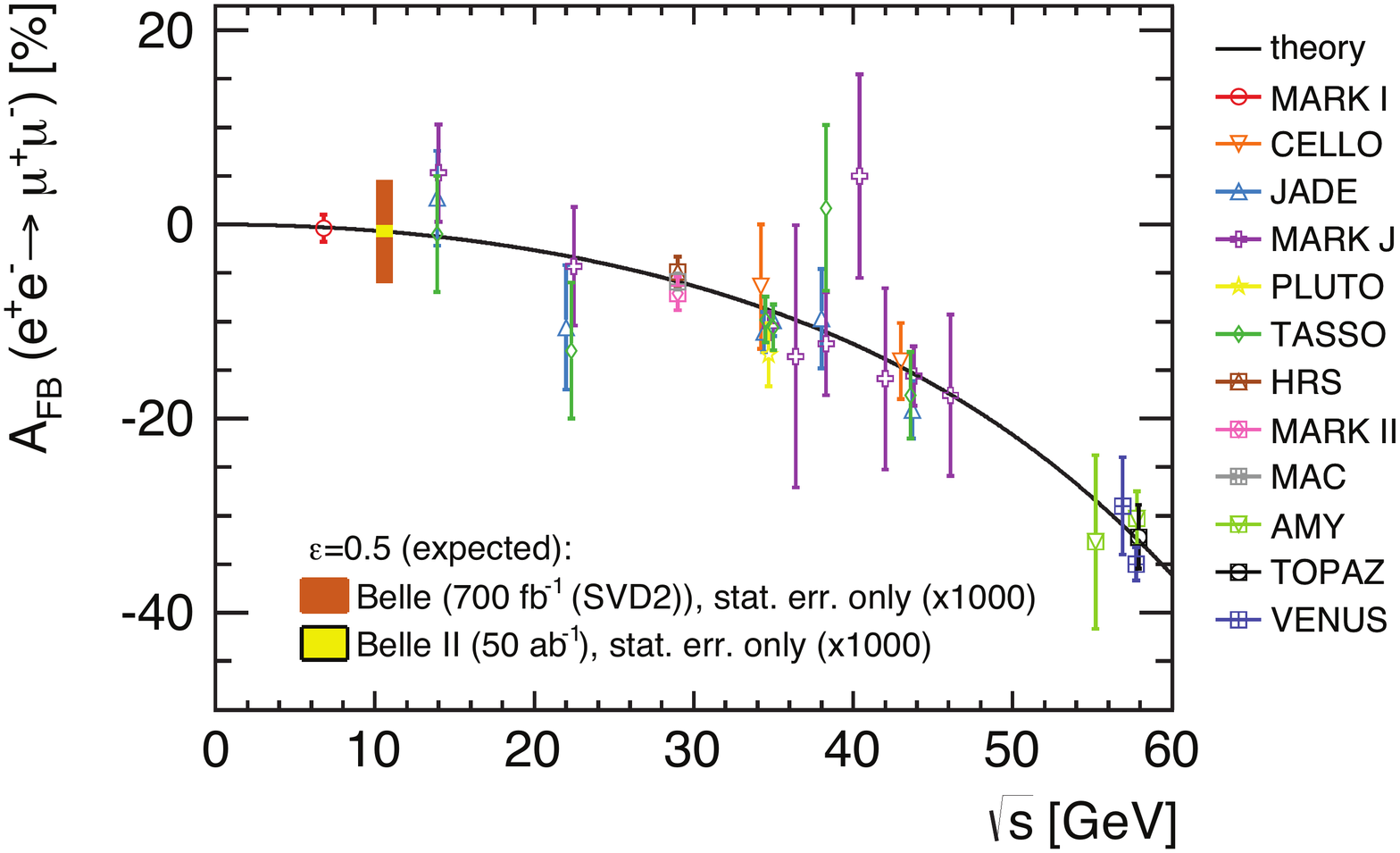}
\vspace*{-0.4cm}
\caption{\label{figferber}
Measurements of $A_{FB}(e^+e^-\to \mu^+\mu^-)$ at different energies corrected
for QED effects by the respective authors (see \cite{PhysRevD.57.5345} 
and references therein), theoretical Standard Model prediction at lowest order and the expected Belle and Belle II statistical uncertainties 
(scaled up by a factor of 1000) at $\sqrt{s} = 10.58$ GeV.
Figure and caption from T. Ferber, in \cite{vanderBij:2014mxa}.
}
\end{figure}

Recently, a quite interesting application of ZFITTER and TOPFIT became relevant at the Belle experiment at KEK in Japan.
In Section 5.14 ``Electroweak physics'' of ``Physics at Super B Factory'' \cite{Aushev:2010bq}, it is worked out that 
Belle II will measure about { $10^9$ $\mu^+\mu^-$ pairs} at { $\sqrt{s}=10.58$ GeV}, with a need of theoretical precision of about 
$10^{-3}$ or even better.
Evidently, to some extent the account of the final state muon mass is needed, ${m_{\mu}^2}/{s}\approx 10^{-4}$. 
This may be exactly controlled by TOPFIT, after some necessary adaptations of the package.
Although the application of ZFITTER was originally excluded at meson factory energies, here it is nevertheless possible due to the 
suppression of contributions from meson resonances in the experimental set-up. 
So, the experiment gives access to electroweak physics at $\sqrt{s} \approx 10$ GeV, from a measurement of
the forward-backward asymmetry $A_{FB}(\mu^+\mu^-)$.
In the usual calculational frames, the weak mixing angle is not accessible, but the measurement is sensitive to the $\rho_Z$ parameter of 
the $Z$ boson.
The $A_{FB}$ will be a single parameter measurement of $\rho_Z$ \cite{Fedorenko:1986hwnew,vanderBij:2014mxa,YashchenkoPRC:2014mod} with high 
accuracy, see figure ~\ref{figferber}. 

\section{Upgrading Monte-Carlo programs for Bhabha scattering  and $\mu^+ \mu^- $ production at the ILC  and at meson 
factories\label{sec-mesonfactoryilc}}

\subsection{Heavy fermion two-loop corrections to Bhabha scattering. Babayaga. \label{subsec-heavy2loop}}
\begin{figure}[t!]
\vspace*{-1.7cm}
\includegraphics[width=8.00cm]{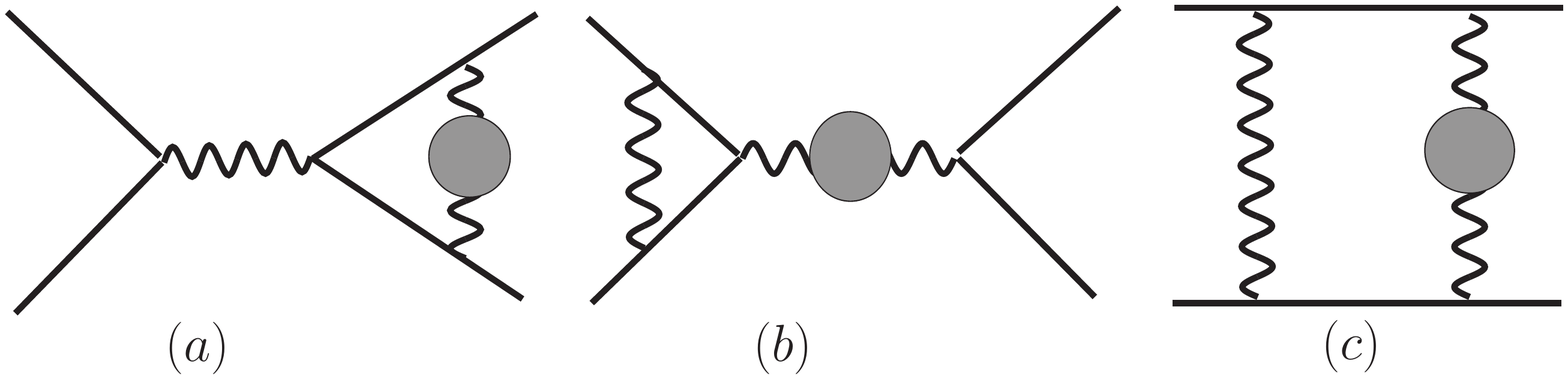}
\vspace*{-2.5cm}
\caption{\label{feynman2loopheavy}The two-loop diagrams with heavy particle corrections to Bhabha scattering.}
 \end{figure}

\begin{figure}[t!]
\center{\includegraphics[width=6.50cm]{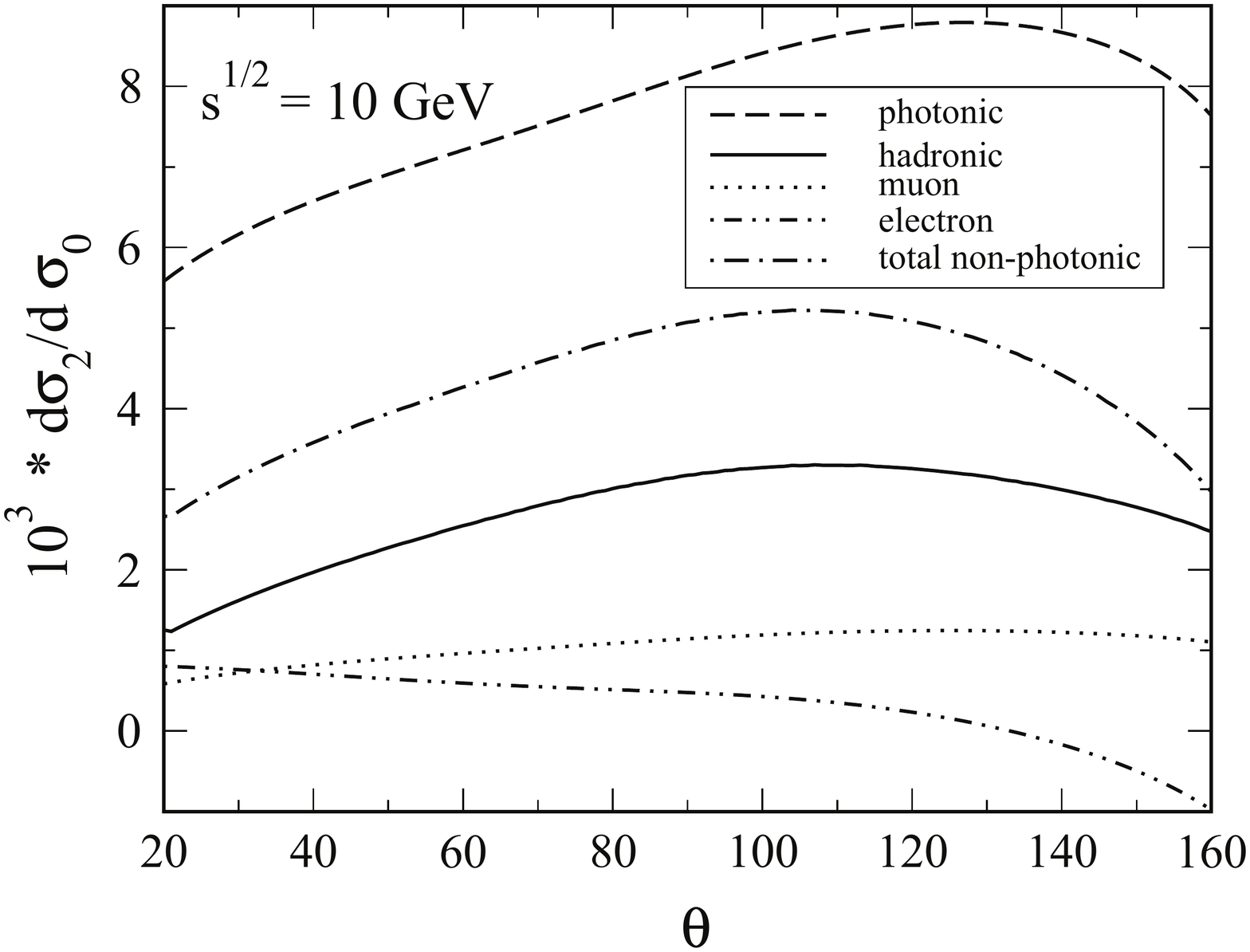}}
\\
\center{\includegraphics[width=6.50cm]{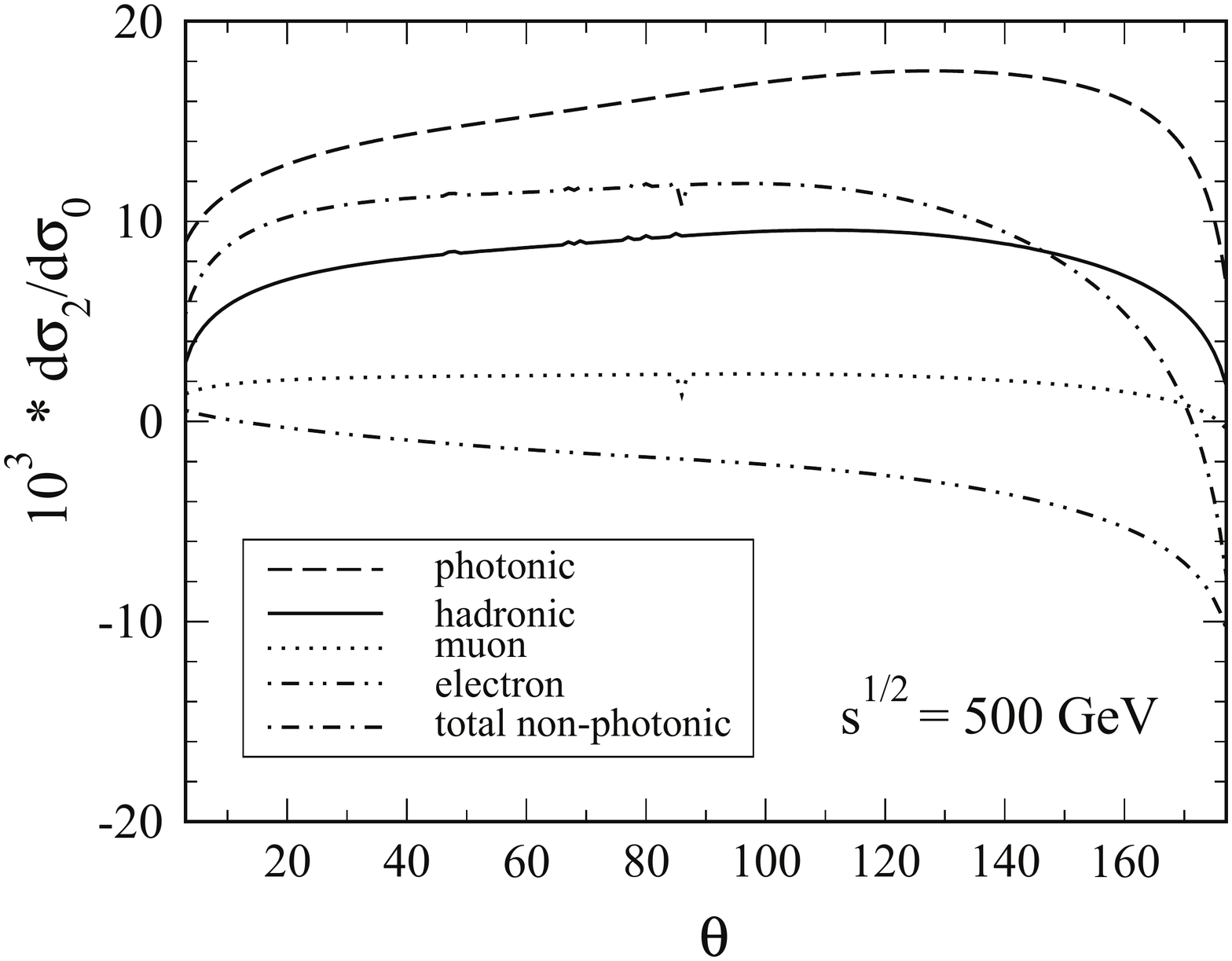}~~~~}
\caption{\label{heavyfermions2loopnumerics}
Massive two-loop corrctions to the differential Bhabha cross section in per mil at two different energies. Figure from \cite{Actis:2007fs}.
}
 \end{figure}

For a variety of scattering processes we need cross section predictions with a two-loop accuracy.
Among them is Bhabha scattering, which is interesting by itself, as a very clean and simple reaction. Any deviation from predictions would 
be a strong indication for New Physics.
Further, small angle Bhabha scattering is important for luminosity determinations at LEP and meson factories, and  both for small and wide 
angle scattering also at the ILC.  
Here, for many applications one may concentrate at QED corrections. An important step was the prediction of the two-loop photonic 
corrections \cite{Penin:2005kf,Penin:2005eh} by relating them to the massless case.
The alternative approach of calculating the virtual two-loop contributions to massive Bhabha scattering proved to be much more involved and 
its proponents have not been succeeded so far. 
We refer to 
\cite{Czakon:2004tg,Czakon:2004wu,Czakon:2004wm,Czakon:2005jd,Czakon:2005gi,Czakon:2006hb,Czakon:2006pa,%
Fleischer:2006ht,Actis:2006dj,Actis:2007gi,Actis:2007pn,Actis:2007fs,Actis:2007zz,Actis:2008sk,Kuhn:2008zs}, and to the recent report 
\cite{Henn:2013woa} on planar integrals.
The difficulty is mainly related to the non-planar double box diagrams.
We come to some aspects and techniques of the evaluation of more complicated Feynman diagrams in section \ref{sec-ambre}.
Much easier, and finally calculated by several groups, are the heavy fermion (and hadronic) two-loop contributions to Bhabha scattering 
shown in figure \ref{feynman2loopheavy} \cite{Actis:2007fs,Actis:2008br,Bonciani:2007eh,Bonciani:2008ep}.
The numerical influence is, both at meson factories and at the ILC, of the order of a per mil, see figure \ref{heavyfermions2loopnumerics}.
The software package related to \cite{Actis:2007fs,Actis:2008br} has been attached to the BabaYaga package
 \cite{Gunia:2011zz,CarloniCalame:2011zq,CarloniCalame:2011aa,Gluza:2012yz} in order to stabilize its numerical 
precision at the NNLO level. 
At the time, there was dedicated scientific activity related to the subject by several groups. We should mention, in addition to the 
extensive list of references quoted in \cite{Actis:2008br}, the references \cite{Actis:2009uq} and 
\cite{Fleischer:1998nb,Bonciani:2003te,Bonciani:2003ai,Bonciani:2003cj,Bonciani:2003hc,Aglietti:2004tq,Bonciani:2004gi,Bonciani:2004qt,
Bonciani:2005im, Aglietti:2007as,Actis:2009zz}.

\subsection{5-point functions in muon pair production.\\ PHOKHARA.}

 \begin{figure} [h]
\hspace*{3mm}\includegraphics[scale=0.35,angle=0]{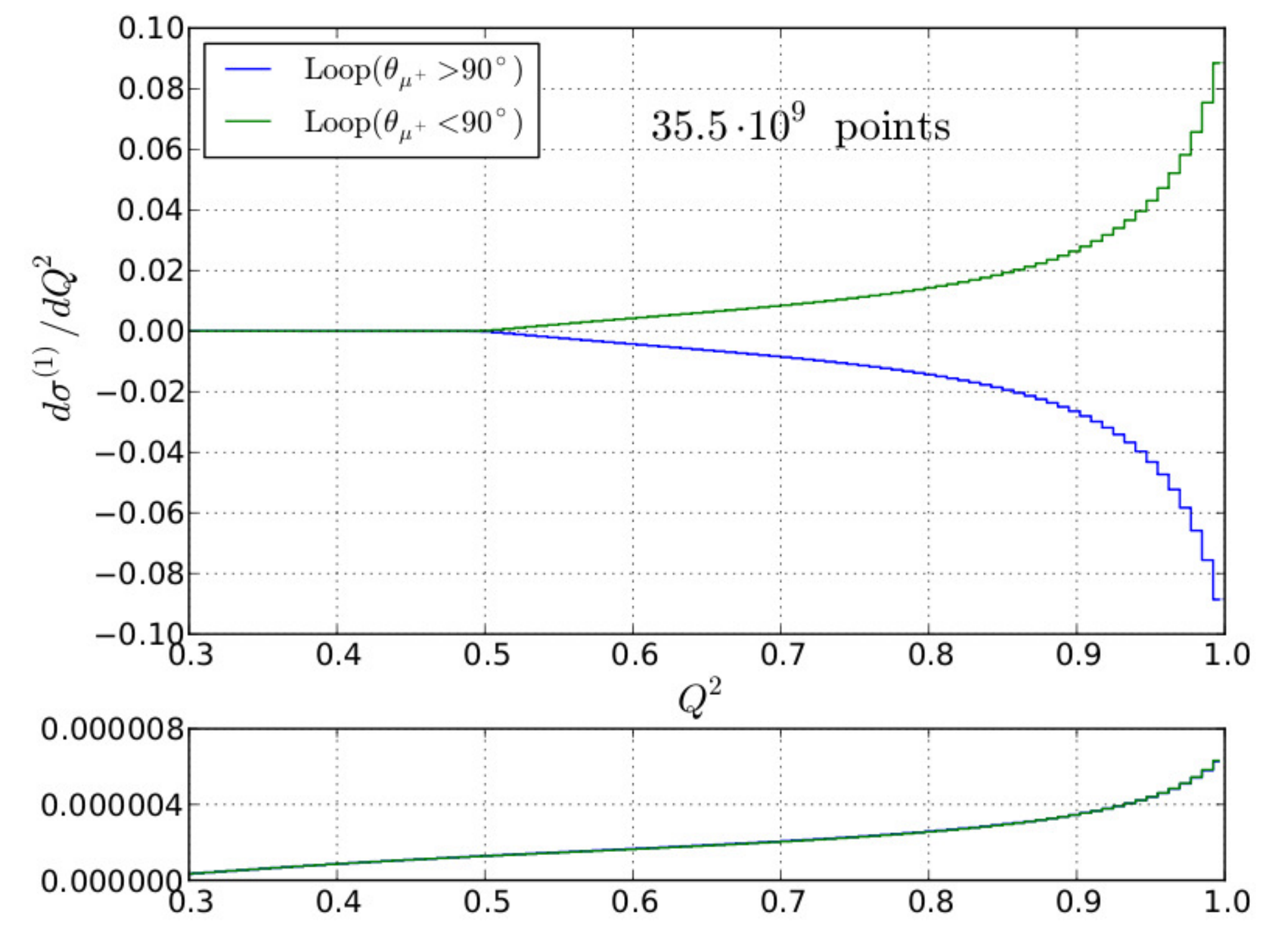} 
\\
\center{\includegraphics[scale=0.25,angle=270]{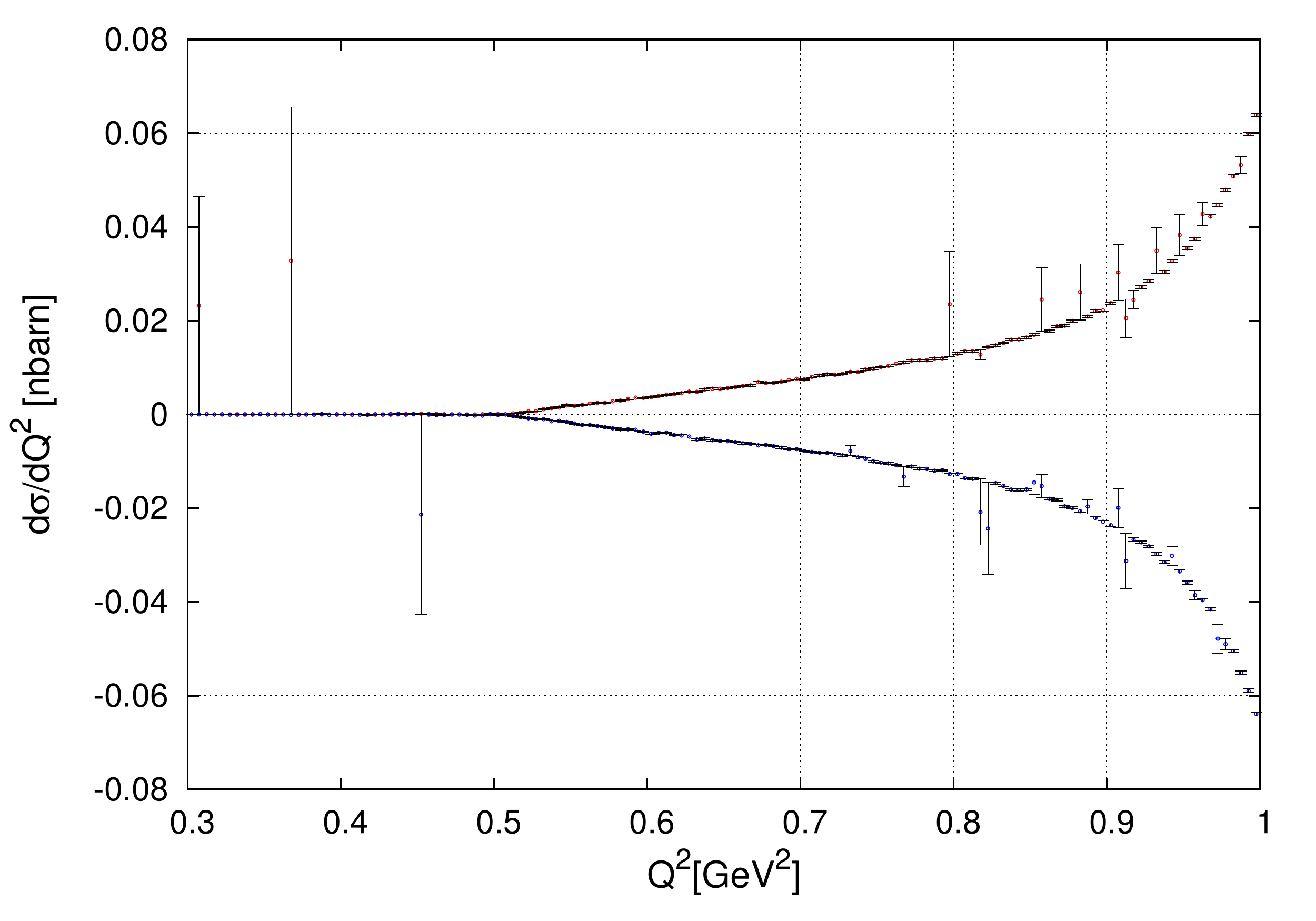}}
\caption{Muon pair distributions including 5-point functions at KLOE
  calculated with PJFry (bottom: absolute error estimate) (a).
  The same calculated without decicated routines to avoid small Gram determinants.
Based on approximately $4\cdot 10^{10}$ ($10^9$) events (b). 
From \cite{Campanario:2013uea}.
\label{fig-acckloe}
}
\end{figure}

\begin{figure}[th!]
\begin{center}
\includegraphics[width=.44\textwidth]{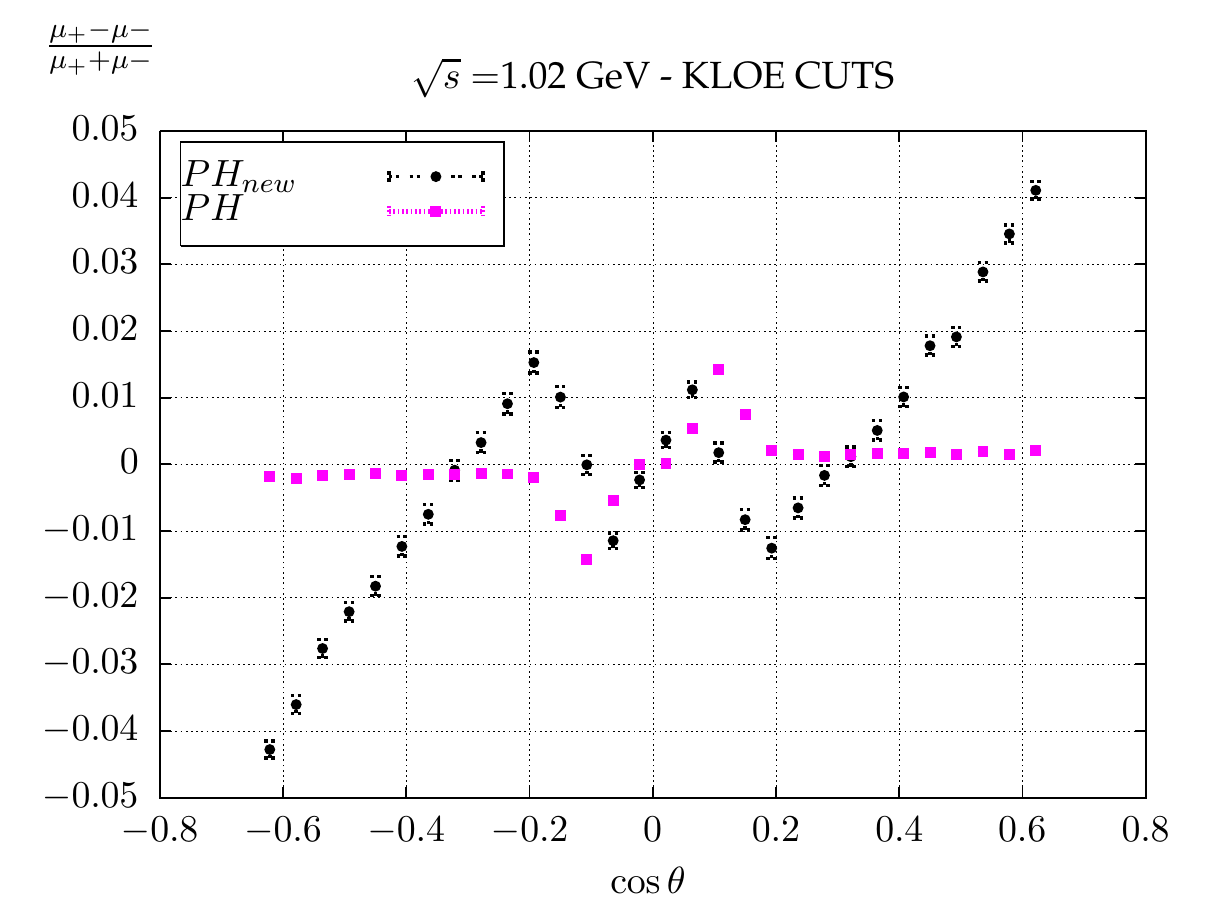}
\includegraphics[width=.44\textwidth]{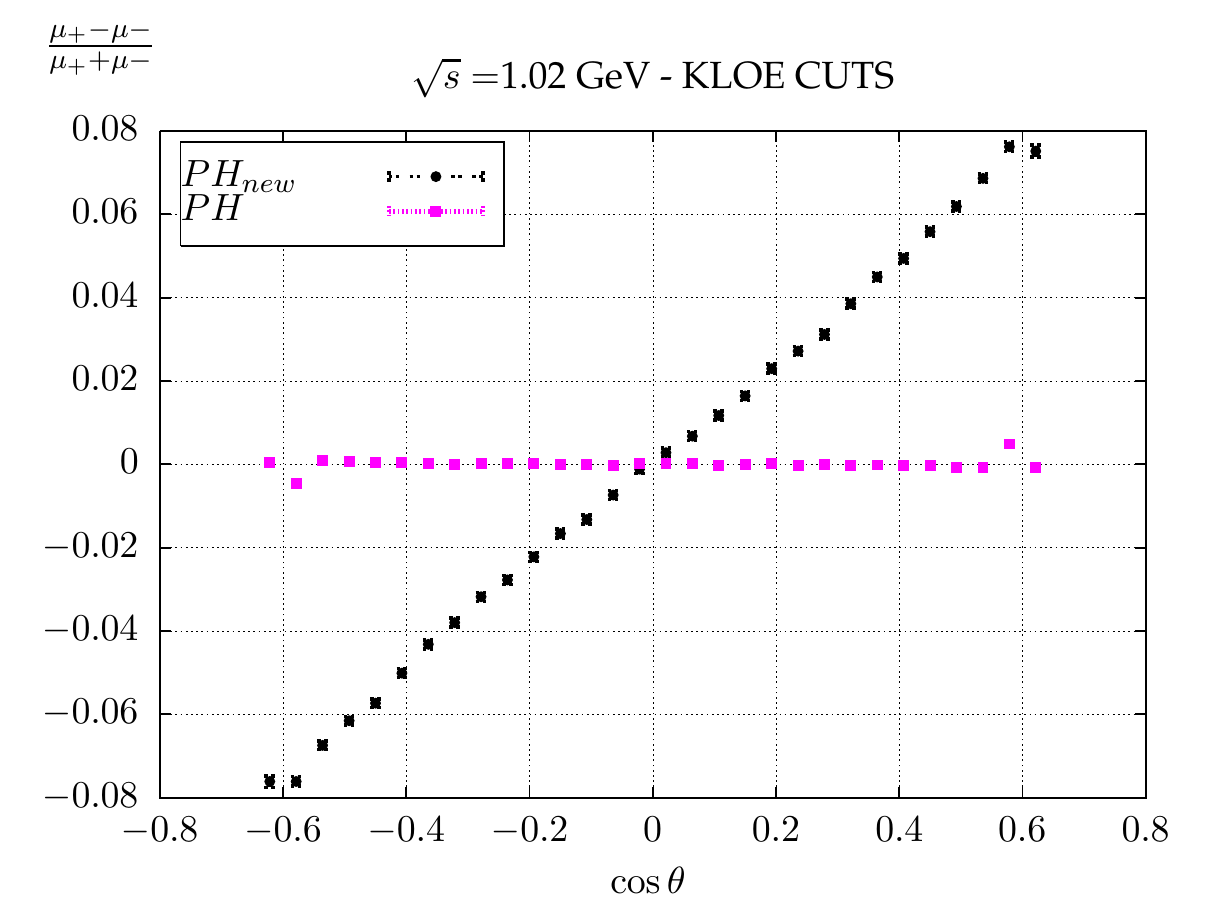}
\end{center}
\caption[]{\label{fig-asymkloe}The asymmetries given by
  PHOKHARA7.0 (denoted as $PH$)
 and PHOKHARA9.0 (denoted as $PH_{new}$). $q^2\in (0.54,0.55)$ - left plot;
 $q^2\in (0.94,0.95)$ - right plot.
From \cite{Campanario:2013uea}.
 }
\end{figure}

Mainly motivated by the need of efficient and numerically stable calculations of higher-point Feynman integrals for the description of 
final states with higher multiplicities at LEP 2 and at the LHC, there was much research activity in recent years on tensor integral 
reduction. 
For a review see e.g. \cite{AlcarazMaestre:2012vp}. 
Section \ref{sec-pjfryolec} is devoted to an advantageous technique, based on dimensional shifts. 
An application of the corresponding software library PJFry is the calculation of the complete QED NLO contributions to 
\bea
\label{mmgg}
e^+e^- \to \mu^+ \mu^- \gamma (\gamma) ,
\eea
which is an important reaction with precision measurements at meson factories. It contains the contributions from five-point 
functions which were not studied before. 
The correct interpretation of the recent (g-2) measurements depends on the reliable knowledge of pion production, and (\ref{mmgg}) is used 
as a normalization for that.
The state-of-the-art Monte Carlo program is PHOKHARA, where the resulting code finally was integrated.
Crucial for stability and efficiency were the inclusion of both the muon and electron masses, and a clever treatment of numerical problems 
related to the appearance of inverse Gram determinants from tensor reduction
\cite{Kajda:2009aa,Gluza:2012yz,Campanario:2013uea,Gunia:2011zz,Yundin-phd:2012oai}.
Figure \ref{fig-acckloe} demonstrates this.   
Figure \ref{fig-asymkloe} shows the improvement from the inclusion of pentagon diagrams on the muon charge asymmetry for KLOE energies for 
two different momentum intervals.
Finally, the numerical effects are below the present experimental concern, what was not evident before.

\section{Tensor reduction for Feynman integrals. PJFry and OLEC.
\label{sec-pjfryolec}}
An efficient approach to the systematic reduction of arbitrary one-loop tensor Feynman integrals to scalar integrals relies on dimensional 
shifts.
Basic ideas  have been formulated in \cite{Davydychev:1991va,Bern:1992em,Tarasov:1996br}, and the approach has been worked 
out in all necessary details and numerical and algebraic tools were created in recent years  
\cite{Fleischer:1999hq,Binoth:2008uq,Diakonidis:2008ij,Fleischer:2010sq,Yundin-phd:2012oai,%
Fleischer:2011zz,Fleischer:2011nt,Diakonidis:2009fx,Fleischer:2011hc,%
Fleischer:2012et,Fleischer:2011zx,Fleischer:2011hc,Fleischer:2011bi,Fleischer:2011nt,Almasy:2013uwa,Fleischer:2012ad,%
Gluza:2012yz}.
The project is competitive with other tools like \cite{Hahn:1998yk,vanOldenborgh:1990yc,Cullen:2011ac,Denner:2014gla}.

\begin{figure}[t!]
\includegraphics[angle=0,height=5.5cm]{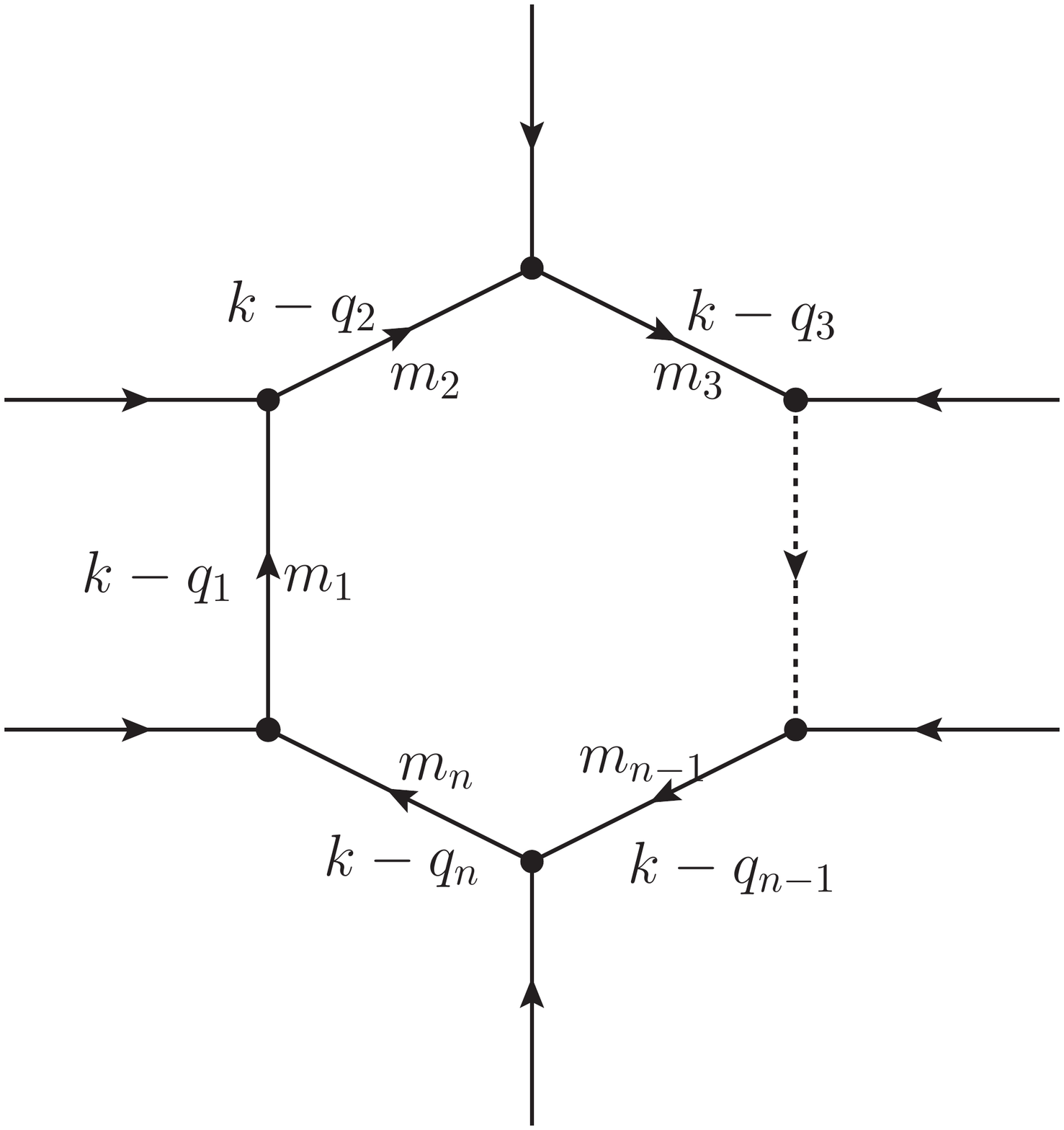} 
\caption{The one-loop $n$ point function.
\label{fig-bueckel}
}
\end{figure}

{ {We consider {$n$-point} tensor integrals of {rank $R$}}:
\begin{equation}
\label{definition}
 I_n^{\mu_1\cdots\mu_R} =  ~~\int \frac{d^d k}{i\pi^{d/2}}~~\frac{\prod_{r=1}^{R}
{k^{\mu_r}}}{\prod_{j=1}^{n}c_j^{\nu_j}}, \nonumber
\end{equation}
where  denominators $c_j = (k-q_j)^2-m_j^2 +i\varepsilon$ have   {\emph{indices} $\nu_j$} and   {\emph{chords}
$q_j$}}.
An example is shown in figure \ref{fig-bueckel}.

The aim is to get for $n > 4$ tensor reductions with:
\begin{itemize}
\item{arbitrary} masses;
\item  { all inverse} pentagon Gram determinants being eliminated; 
\item  {treatment of} full kinematics, also with vanishing sub-diagram Gram determinants;
\item  higher $n$ point functions, {\bfseries{ {$n \geq 6$}} } \cite{Fleischer:2011hc};
\item   as an option: multiple sums over tensor coefficients made efficient 
by {{contracting with external momenta }} \cite{Fleischer:2011nt}.
\end{itemize}
Tensor integrals in terms of { scalar integrals} ${I_{n,i\cdots}^{[d+]^k}}$
in higher dimensions, {$D=d+2l~=~4-2{\epsilon},~6-2{\epsilon}, \cdots$} 
were derived in \cite{Davydychev:1991va}, see also \cite{Fleischer:1999hq}.
With
{$n_{ij}={\nu}_{ij}=1+{\delta}_{ij},n_{ijk}= {\nu}_{ij}{\nu}_{ijk},
{\nu}_{ijk}=1+{\delta}_{ik}+{\delta}_{jk} $}, one gets for a tensor of rank four:
\begin{eqnarray}
I_{n}^{\mu\, \nu\, \lambda\, \rho} &=& 
\int ^{d} k^{\mu} \, k^{\nu} \,  k^{\lambda} \, k^{\rho} \, \prod_{r=1}^{n} \, {c_r^{-1}} 
\nn\\
&=&    \sum_{i,j,k,l=1}^{n} \, q_i^{\mu}\, q_j^{\nu}\, q_k^{\lambda} 
 \, q_l^{\rho}\,  n_{ijkl} \,  \, {I_{n,ijkl}^{[d+]^4}}  
\nonumber\\
\nn\\
&&    -~\frac{1}{2} \sum_{i,j=1}^{n} {g^{[\mu \nu}} q_i^{\lambda} q_j^{\rho]}
\, n_{ij} {I_{n,ij}^{[d+]^3}}
\nn\\
&& +~\frac{1}{4} {g^{[\mu \nu} g^{\lambda \rho]}} {I_{n}^{[d+]^2}} .
\end{eqnarray}

\begin{figure}[t]
\begin{center}
{\includegraphics[width=.27\textwidth]{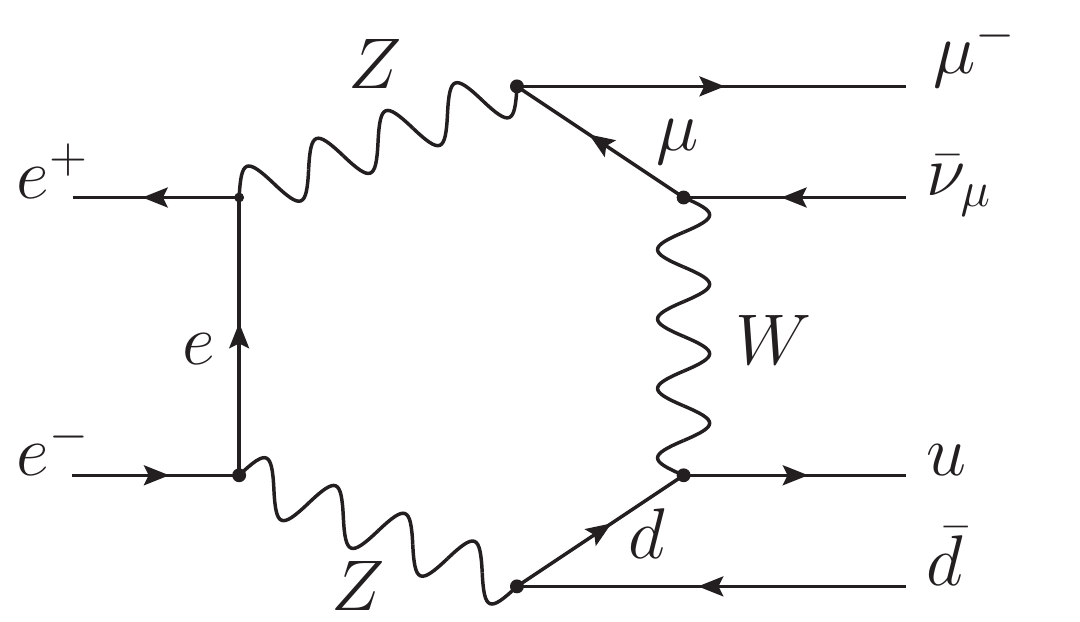}}
\\
{\includegraphics[width=.27\textwidth]{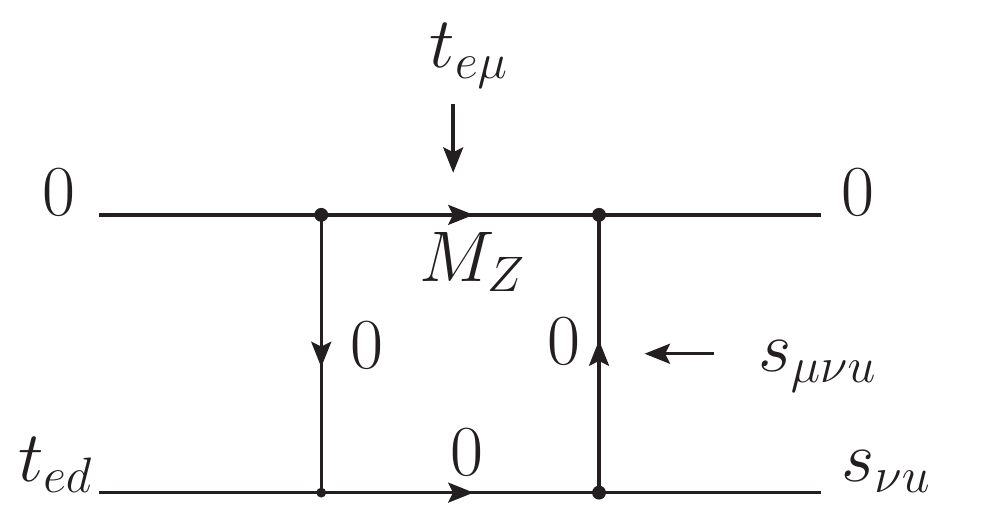}}
\end{center}
\caption[]{%
{
{A six-point topology  leading to a four-point function with realistically vanishing Gram determinant. Example from \cite{Denner:THHH2009}. 
}\label{fig-4-and-6-point}
}
}
\end{figure}

In order to use a publicly available library of scalar functions, one has to lower the dimensions $D = [d+]^n = d+2n$ on the right hand 
side of the 
reductions to $d = 4-2\epsilon$.
This may be done by Tarasov's dimensional recurrences \cite{Tarasov:1996br,Fleischer:1999hq}, based on the notion of signed minors 
\cite{Melrose:1965kb}:
\begin{equation}
{\small 
\label{eq:RR1}
 \nu_j    \bigl( {\bf j^+} I_5^{{ [d+]}} \bigr)
=
\left[  - {{j \choose 0}_5} +\sum_{k=1}^{5} {j \choose k}_5
   {{\bf k^-} }\right]   \frac{I_5}{\left(\right)_5} ,
}
\end{equation}
\begin{equation} 
\label{eq:RR2}
  (d-\sum_{i=1}^{5}\nu_i+1)      I_5^{[d+]}
=
  \left[ {{0 \choose 0}_5}
 - \sum_{k=1}^5 {0 \choose k}_5 {{\bf k^-}} \right]   
\frac{I_5}{\left(\right)_5},
\end{equation}
with appearance of a dimensional shift and of an inverse Gram determinant $\left( \right)_5$,
and the integration-by-parts relation \cite{Chetyrkin:1981qh}:
\begin{equation} 
{\small 
 \nu_j   {\bf j^+} I_5
=
  \sum^{5}_{k=1} {0j\choose 0k}_5
\left[ d - \sum_{i=1}^{5} \nu_i(   {{\bf k^-}}  { {\bf i^+}}+1)
             \right]   
\frac{I_5}{ {0\choose 0}_5 } .
}
\end{equation}
The operators   ${\bf  i^{\pm}, j^{\pm}, k^{\pm} }$ shift 
the indices   {$\nu_i, \nu_j, \nu_k$} by $\pm 1$.
As an example, figure \ref{fig-4-and-6-point} shows a six-point function with a kinematically critical four-point function in the reduction.

\begin{figure}[t]
\center{ \includegraphics[angle=0,width=.4\textwidth]{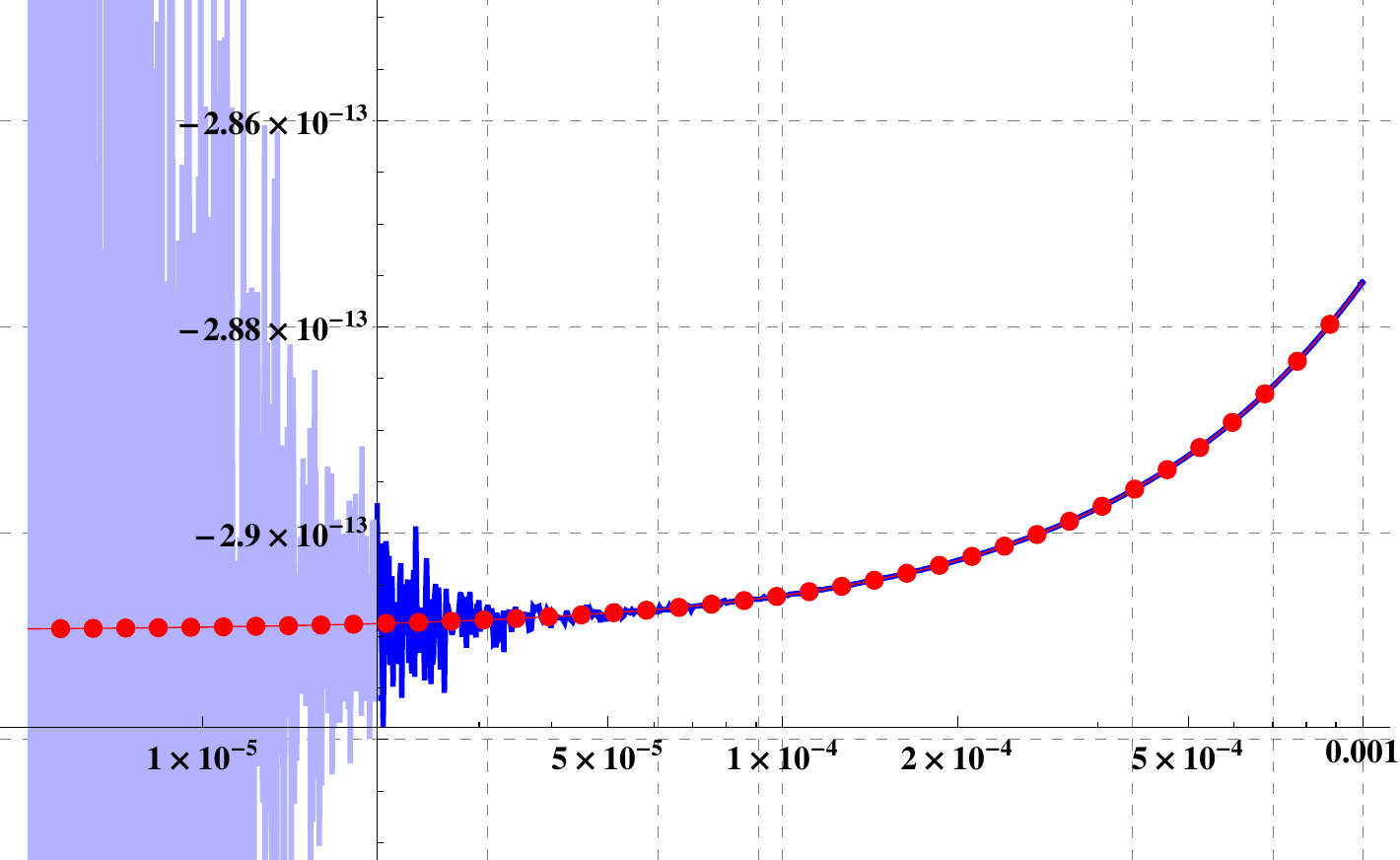} }
\caption{
$E_{3333}$ coefficient in small Gram region $(x\to 0)$, from \cite{yundin-lhcphenonet2011}.
    Comparison of regular (blue) and expansion (red) formulae.
$x=0$: $E_{3333}$(0, 0, $-6{\times}10^4$, 0, 0,
             $10^4$, $-3.5{\times}10^4$, $2{\times}10^4$, $-4{\times}10^4$, $1.5{\times}10^4$,
             0, 6550, 0, 0, 8315).
\label{fig-pjfry-example}
}
\end{figure}

We may identify our tensor coefficients in  LoopTools/FF notations \cite{Hahn:1998yk,vanOldenborgh:1990yc}, e.g.
$     I_{4,222}^{[d+]^3} = D_{111}$, and similarly $ I_{4,2222}^{[d+]^4} = D_{1111}$.      
Figure \ref{fig-pjfry-example} shows the numerical stabilization near a phase space point with vanishing Gram determinant, which is 
reached by an expansion of the higher-point functions in a series of higher-dimensional three-point functions (so avoiding inverse Grams) 
with their subsequent dimensional recurrence, improved finally by a Pade approximation \cite{Fleischer:2010sq}. 

The C++ package PJFry for tensor reduction is due to V. Yundin and is available as open-source software \cite{pjfry-project}.
It has been used for a variety of cross section calculations for LHC and meson factories.
The C++ package OLEC \cite{olec-project-09} for the replacement of tensor reduction by tensor contractions is due to J. Gluza, M. Gluza, I. 
Dubovyk. Its basics have been worked out also by A. Almasy, J. Fleischer, T. Riemann \cite{almasy-reductions0-f99,Almasy:2013uwa}. It is 
not yet published. 

\section{\label{sec-ambre}Calculation of master integrals with Mellin-Barnes representations. AMBRE.}
\begin{figure}[t]
\vspace*{-3cm}
\center{  \includegraphics[width=6.0cm]{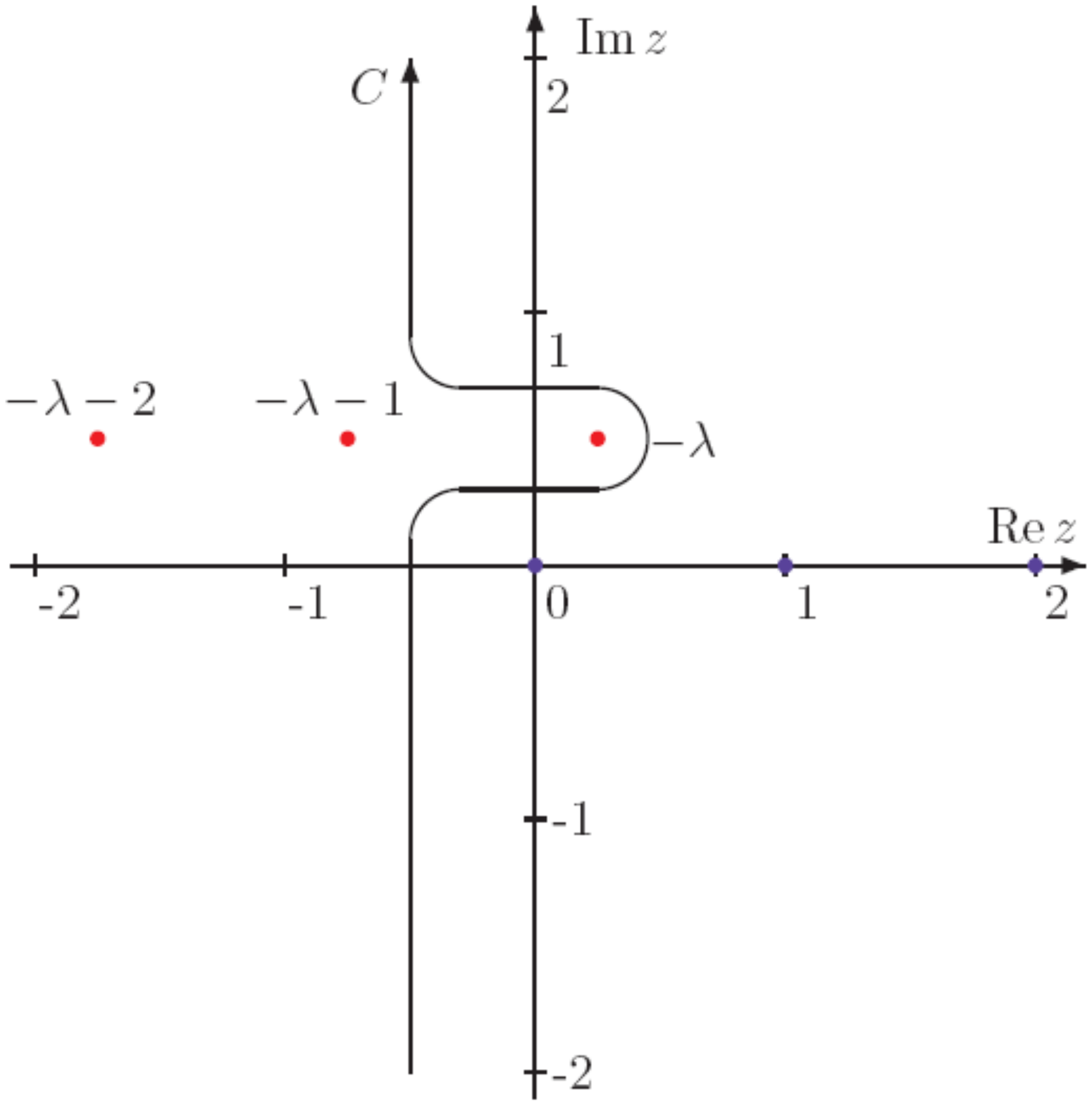}  }
\caption{\label{fig-mbpath}The integration paths of the multi-dimensional Mellin-Barnes integral have to be chosen properly and to be 
closed to the left or to the right at infinity.}
\end{figure}

There are several powerful techniques for the evaluation of higher order Feynman integrals.
We mentioned already the reduction to (an algebraic system of) simpler integrals by recurrence relations.
A similar idea is followed by solving (systems of) difference or differential equations. 
Single Feynman integrals 
may be solved by a multiple Feynman parameter integral representation for the propagators,
\bea
 \frac{1}{D_1^{n_1} D_2^{n_2} \ldots D_N^{n_N}} 
= 
\frac{\Gamma(n_1+ \ldots + n_N)}{\Gamma(n_1) \ldots \Gamma(n_N)} 
 \int \limits_{0}^{1} d x_1 
\nonumber \\
\ldots \int \limits_{0}^{1} d x_N 
 \frac{x_1^{n_1-1} \ldots x_N^{n_N-1} \delta(1 - x_1 - \ldots - x_N)}
      {(x_1 D_1 + \ldots + x_N D_N)^{N_{\nu}}} ,
\eea
where $N_{\nu} = n_1+ \ldots + n_N$.
After performing the momentum integrations in (\ref{eq-2}), the $x$-parameters are left:
\bea
 G(X) =  
 \frac{(-1)^{N_{\nu}} \Gamma\left(N_{\nu}-\frac{d}{2}L\right)}
 {\prod \limits_{i=1}^{N}\Gamma(n_i)}
 \int \prod \limits_{j=1}^N dx_j ~ x_j^{n_j-1}
\nonumber\\
 \times ~\delta(1-\sum_{i=1}^N x_i)
 \frac{ {U(x)}^{N_{\nu}-d(L+1)/2}}{ {F(x)}^{N_{\nu}-dL/2}} .
\eea 
 The functions { {$U$}} and { {$F$}} are called graph or Symanzik polynomials. 
Sector decomposition techniques allow to isolate endpoint singularities related to infrared poles.
Another technique is the replacement of sums by products of $x$-monomials in the Symanzik polynomials   {$U$} and  {$F$} by a 
sequence of Mellin-Barnes (MB-) integrals \cite{zbMATH02640947} with well-defined integration paths in the complex plane, see figure 
\ref{fig-mbpath}. 
\bea \label{eq-18}
 \frac{1}{({ {A_1}}+\ldots+{ {A_n}})^{\lambda}} = \frac{1}{\Gamma(\lambda)}\frac{1}{(2 \pi i)^{n-1}}
 \nonumber \\ \nonumber 
\times ~\int_{-i \infty}^{i \infty} dz_1 
\ldots dz_{n-1} 
 \prod \limits_{i=1}^{n-1} { {A_i}}^{z_i} ~
 { {A_n}}^{-\lambda - z_1 - \ldots - z_{n-1}}
\nonumber \\
 \times~  \prod \limits_{i=1}^{n-1} \Gamma(- z_i) ~
 \Gamma(\lambda + z_1 + \ldots + z_{n-1}) .
\eea
 
In recent years, several software packages in MATHEMATICA have been developed 
for the calculation of (\ref{eq-18}) and are widely used
\cite{mbtools,Czakon:2005rk,Smirnov:2009up,Gluza:2007rt,Gluza:2010rn,Gluza:2007bd}.
The Mellin-Barnes approach has certain limitations, partly resting in its basic features. 
We like to mention the dimensionality. 
The dimensionality of the sector decomposition 
\cite{Bogner:2007cr,Heinrich:2008si,Kaneko:2009qx,Gluza:2010rn,Borowka:2012yc,Borowka:2013cma,Smirnov:2013eza} is 
essentially the number of Feynman parameters, while in the 
MB-approach it depends on the complexity of the problem and may be quite high.
Further, only recently there was substantial progress for the treatment of non-planar topologies with  AMBRE 
\cite{Bielas:2013rja,Blumlein:2014maa}, while numerics in the Minkowskian region is not yet implemented.
An MB-treatment of mixed virtual and real IR-singularities has been proposed in \cite{Gluza:2008tk}.
An advantageous feature of MB-integrals is the potential to find analytical solutions.
This may be tackled by deriving infinite series over residues with the Cauchy theorem \cite{Riemann:2012linznoit}.
Some software is under development \cite{Blumlein:2014maa}.
The bottleneck is the need to sum up the multiple series. 
While the algorithms of e.g. MATHEMATICA are limited, there are dedicated approaches like SUMMER \cite{Vermaseren:1998uu} 
and
XSUMMER \cite{Moch:2005uc}, for massless problems of a not too high dimensionality.
Both packages are written in FORM \cite{Vermaseren:2000nd}.
Presently,we are exploring the potential of the package family of the RISK (Linz) group around the MATHEMATICA package SIGMA 
\cite{SIG1,SIG2}.  
For automation we have to implement several steps:
\begin{enumerate}
\item {Determine if the topology} is planar or non-planar; PlanarityTest \cite{Bielas:2013rja}.
\item {Construct   MB representations}; AMBRE, MB, and related packages 
\cite{mbtools,Czakon:2005rk,Smirnov:2009up,Gluza:2007rt,Gluza:2010rn}. 
\item Change the MB-integrals into nested sums;  MBsums package \cite{Blumlein:2014maa}.
\item {Try to perform  the multiple sums} analytically \cite{Blumlein:2014maa,Ablinger:2014wca}.
\item  {Accept Minkowskian} kinematics. 
\end{enumerate}
As mentioned, there are intrinsic limitations to the MB-approach, due to the number of loops, the number of scales, the number 
of legs, and last but not least the complexity of a given integral.

In a practical study, one will  combine several methods in order to calculate all the necessary Feynman integrals.
One or the other application \cite{Actis:2006dj,Czakon:2006pa} was mentioned already in the foregoing sections.
As an interesting state-of-the-art problem we mention here the calculation of the complete two-loop electroweak vertex 
form factors of the $Z$ boson 
where one has to determine several hundreds of Minkowskian Feynman integrals of different dimensionality with up to four different 
dimensionless scales (arising from $s, M_Z, M_W, M_H, m_t$) and with an accuracy of several numerical digits, typically about five. 
Alternatively, one might try to  determine a minimal set of master integrals, but this 
raises other non-trivial problems.
Concerning mulit-scale problems, notably arising in electroweak mulit-loop calculations, the numerical approach has benn proved to be 
powerful. We can quote here only a selection of relevant articles 
\cite{Passarino:2001wv,Passarino:2001jd,Ferroglia:2002mz,Becker:2012bi,Ferroglia:2003yj,Actis:2004bp,Anastasiou:2005cb,Pozzorini:2005ff,
Passarino:2006gv, Actis:2006xf,Anastasiou:2007qb,Freitas:2010nx,Ellis:2011cr,Carter:2011uza,Becker:2012nk,Freitas:2012iu}.  
\section{The role of software in the dissemination of scientific results\label{sec-dissemination}
}
The dissemination of the results of theoretical research in elementary particle physics is an essential element of any single project.
We mentioned in the introduction that the publication of an article in a peer-reviewed scientific journal is not sufficient in view of the 
complexity of contemporary calculations.
Additionally, one has to deliver some piece of software code ready for use by third parties, notably experimentalists. It has to be 
sufficiently robust to be safely used by non-specialists and during longer periods,  maybe even decades.
Further, it is often needed to deliver some direct individual support by debugging, new adaptations, etc.
Some aspects of long-term support have been described for the example of the ZFITTER project \cite{Akhundov:2014era}, see also section 
\ref{sec-zfitter}.  

There are the rapidly rising  opportunities of interconnections by the internet.
More and more often they are used anonymously, essentially because the number of researchers is also rapidly rising. Just compare the 
number of scientists per experiment at LEP (few hundreds) and at LHC (three thousand).  
This raises questions related to the foundations of academic basic research like proper quotation, and related to copyright law, like 
licence problems.

There is an ongoing discussion in a broader scientific community in Germany on software use in 
scientific cooperation.   
We quote from a statement by the Ombudsman for Science in Germany, formulated with 
respect to software use in high energy physics (3 July 2012) \cite{ombuds:2012}:

{\it 
``The proper legal treatment of such software, in the field of tension of the rules of good scientific practice, has not been the subject 
of rule making until now (if I see right) \ldots''
}

A similar understanding was expressed by the Editor-in-Chief of ``European Physical Journal C'' (26 January 2012) 
\cite{editorEPJC:2012}:

{\it 
``We note that a subtlety may remain in the question as of what "scientific usage of the code" includes in the broader sense, namely if it 
is restricted to using the code as-is, or if copying and altering the original code is also permitted. Here we refer to the common practice 
of e.g. using Monte Carlo generator code
by a large number of scientists who, as we observe, not only run that original code, but alter and copy parts of it according to their 
specific (scientific) needs.
Such Monte Carlo codes exist, in a wide variety, under similar or identical license terms, as Open Source software \ldots''
}

These quotes might be contrasted by the scientific practice in our international community.  
Let us refer to the write-up of the results of the ACAT (May 2013, Beijing) round table discussion on ``Open-source, knowledge sharing and 
scientific collaboration'' \cite{Carminati:2014yra,Carminati:2014yya} where many interesting reflections and ideas, together with facts of 
life, where communicated.
The abstract summarizes: 

{\it 
``Although the discussion was, in part, controversial, the participants agreed unanimously on several basic issues in software sharing:
\begin{itemize}
 \item The importance of having various licensing models in academic research;
\item The basic value of proper recognition and attribution of intellectual property, including scientific software;
\item The user respect for the conditions of use, including licence statements, as formulated by the author.''
\end{itemize}
}

In the US, there is governmental interest in the sensible topic.
On June 3, 2011, the  ``Report of the HEPAP Sub-Committee on the Dissemination 
of Research Results'' \cite{Artuso:2011cqa} was published which answers a request by the Director of the US DOE Office of 
Science  to summarize the current practices of researchers funded by the Office of High Energy
Physics (OHEP) for disseminating their results.
We quote from there:

{\it  
``Although not technically ''digital data``, it's important to note that some theoretical
research produces results besides the published articles. Examples include simulation
programs (e.g. lattice gauge theory simulations like USQCD or MILC and Monte Carlo
simulation programs like PYTHIA, HERWIG, SHERPA, or ALPGEN), computation programs (e.g. MCFM or MadGraph), and
global fits to a large corpus of data (e.g. CTEQ, ZFITTER, or CKMFITTER). Typically the computer code itself is disseminated in an open 
access manner via the internet. The release of the computer code is usually
accompanied by a publication in a peer reviewed journal describing the functionality of
the code and, if relevant, specific results obtained using the code \ldots
The Version of Record is taken to be the latest version available from the relevant URL,
which also provides additional functionality by providing versioning, documenting the
relevant differences among versions, producing a User’s Manual, and referencing the
related articles in peer reviewed journals and/or posted on the arXiv. The long-term
stewardship of these results is provided by the collaborations themselves via their web
pages.''
}

There is a high conformity between statements in the ACAT round table summary  \cite{Carminati:2014yra,Carminati:2014yya} and in the 
HEPAP Sub-Committee report \cite{Artuso:2011cqa}, although there was absolutely no interaction between their authors. 

These quotes might stimulate further thinking on the subject.

\section*{Acknowledgements}

We remember thankful the cooperation with  Prof. Dr. Jochem Burkhard Fleischer 
(17 December 1937 - 1 April 
2013).\footnote{\href{http://www-zeuthen.desy.de/~riemann/FleischerJ}{http://www-zeuthen.desy.de/$\sim$riemann/FleischerJ}}

\bigskip

We profited essentially from several long-term cooperations.
Instead of listing here a list of names, we would like to refer to the exhaustive list of references where we give full account to our
co-authors. 

\bigskip

Most of the work presented would not have been possible without the support by DFG Sonderforschungsbereich Transregio 9, 
Computergest{\"u}tzte
Theoretische Teilchenphysik.
We also acknowledge support by the Research Executive Agency (REA) of the European 
 Union under the Grant Agreement number PITN-GA-2010-264564 (LHCPhenoNet) and by the Polish National Center of Science (NCN) under the 
Grant Agreement
number DEC-2013/11/B/ST2/04023.


\bibliographystyle{elsarticle-num}
\bibliography{2loops}

\end{document}